%% file: paper_arxiv.tex
\documentclass[12pt]{article}

\include{_preamble}

\AtEndDocument{\refstepcounter{theorem}\label{finalthm}}
\AtEndDocument{\refstepcounter{equation}\label{finaleq}}
\title{\titlepaper}

\begin{document}
\maketitle

\begin{abstract}
Interventional effects have been proposed as a solution to the
unidentifiability of natural (in)direct effects under
mediator-outcome confounders affected by the exposure. Such
confounders are an intrinsic characteristic of studies with
time-varying exposures and mediators, yet the generalization of the
interventional effect framework to the time-varying case has
received little attention in the literature. We present an
identification result for interventional effects in a general
longitudinal data structure that allows flexibility in the
specification of treatment-outcome, treatment-mediator, and
mediator-outcome relationships. Identification is achieved under the
standard no-unmeasured-confounders and positivity assumptions. We
also present a theoretical and computational study of the properties
of the identifying functional based on the efficient influence
function (EIF). We use the EIF to propose a sequential regression
estimation algorithm that yields doubly robust,
$\sqrt{n}$-consistent, asymptotically Gaussian, and efficient
estimators under slow convergence rates for the regression
algorithms used. This allows the use of flexible machine learning
for regression while permitting uncertainty quantification through
confidence intervals and p-values. A free and open source \texttt{R}
package implementing our proposed estimators is made available on
GitHub. We apply the proposed estimator to an application from a
comparative effectiveness trial of two medications for opioid use
disorder. In the application, we estimate the extent to which
differences between the two treatments' on subsequent risk of opioid
use is mediated by craving symptoms.
\end{abstract}

\section{Introduction}

Mediation analyses have a long history in the statistical sciences
and have more recently been proposed in the causal inference literature
as a way of understanding the mechanisms through which effects operate
\citep{vanderweele2009mediation}. For example, recent mediation
analyses have helped to uncover the types of immune response that
COVID-19 vaccines trigger in order to prevent disease
\citep{gilbert2021immune}.

Multiple methods have been developed for mediation analysis in the
setting of a mediator measured at a single time point, using a
counterfactual framework. \cite{RobinsGreenland92} and \cite{Pearl01}
defined and established conditions for identifiability for so-called
\textit{natural direct effect} and \textit{natural indirect effects},
which measure the difference in counterfactual outcomes under certain
simultaneous interventions on the treatment and mediator, and
decompose the total average treatment effect into direct and indirect
effects. Though the definition of the natural (in)direct effects is
scientifically interesting, their identification requires a so-called
\textit{cross-world} counterfactual independence assumption, which
generally precludes the presence of (measured or unmeasured)
intermediate confounders of the mediator-outcome relation affected by
treatment \citep{andrews2020insights}. This restriction limits the
applicability of natural direct and indirect effects in practice, as
intermediate confounders are expected to be present in many
applications, for example when the effect of an intervention operates
through adherence \citep{rudolph2021helped}.

Several methods have been proposed to do away with the cross-world
independence assumption and/or the assumption of no intermediate
confounders. For example,
\cite{robins2010alternative,tchetgen2014bounds,miles2015partial}
present partial identification methods, \cite{robins2010alternative}
introduced so-called \textit{separable effects}, and
\cite{diaz2020causal,hejazi2020nonparametric} propose the definition of the effects in terms
of stochastic interventions on the exposure. In this paper we focus on
the approach of \cite{petersen2006estimation, van2008direct,
  vanderweele2014effect} and \cite{vansteelandt2017interventional},
who propose to define mediation effects in terms of contrasts between
counterfactuals in hypothetical worlds in which the treatment is set
to some value deterministically, whereas the mediator is drawn
stochastically from its counterfactual distribution under
interventions on treatment. Efficient non-parametric estimators that
leverage machine learning to alleviate model misspecification bias
have been recently proposed for these parameters
\citep{diaz2021nonparametric}. These effects have been called
\textit{interventional} effects \citep{vanderweele2017mediation}, a name we adopt in this paper.

While the definition of interventional effects has allowed researchers
to make progress in settings with intermediate confounding, several
limitations remain, especially in the setting of variables measured at
multiple points in time.  In a seminal paper in longitudinal mediation
analysis, \cite{vanderweele2017mediation} present a non-parametric
identification formula for interventional effects in the case of
mediators and treatments measured longitudinally, assuming that the
time-dependent covariates are measured either before or after the
mediator, but not both. The authors propose estimation methods that
rely on the unlikely ability to correctly specify parametric models
for the distribution of the unobservable counterfactual outcomes.
\cite{zheng2017longitudinal} propose similar interventional effects
where the mediator is drawn from its counterfactual distribution
conditional on all the past, and develop non-parametric efficient
estimators that rely on data-adaptive regression to alleviate model
misspecification bias. However, the ``direct effect'' defined in
\citet{zheng2017longitudinal} does not capture the pathway from
treatment through intermediate confounder to outcome,
and is thus not a direct effect in the
sense that we are interested in this paper. In a similar vein,
\cite{vansteelandt2019mediation} develop methods for survival analysis
and treatment at a single time point. They define the indirect effect
as an effect that measures only paths where the treatment directly
affects the mediator (and not paths where there is an indirect effect
of the treatment on the mediator), and show that such effects are
identified even in the presence of longitudinal confounding of the
mediator-outcome relation affected by treatment.
\cite{bind2016causal} propose effects for longitudinal treatments and
mediators, but does not allow for intermediate confounders and
requires the cross-world counterfactual
assumption. \cite{mittinty2020longitudinal} propose marginal
structural models for longitudinally measured mediators under
treatment at a single time point and no loss-to-follow-up, and propose
estimators that are based on parametric marginal structural models on the
counterfactual outcomes.

In this paper we develop a general longitudinal causal mediation approach that
fills several gaps from the above literature. The method we propose
satisfies the following: (i) the direct effect is defined
in terms of the effects operating through all the pathways that do not
include the mediator at any time point, (ii) allow for longitudinally
measured mediators and treatments, with confounders possibly measured
before and after treatment, (iii) allow for the use of data-adaptive
regression to alleviate concerns of model misspecification bias, and
(iv) allow the construction of efficient estimators and computation of
valid standard errors and confidence intervals, even under the use of
data-adaptive regression. A limitation of prior work remains: our proposed
methods can only handle categorical mediators, and the computational
complexity increases with the number of categories. 

The remainder of
the paper is organized as follows. In \sec\ref{sec:notation} we
introduce the parameters of interest as well as the identification
result, in \sec\ref{sec:eff} we discuss efficiency theory for the
interventional mediation functional, presenting estimating equations
and the efficiency bound in the non-parametric model, in
\sec\ref{sec:estima} we discuss the proposed estimator as well as its
asymptotic properties, and finally in \sec\ref{sec:aplica} we present
the results of an illustrative study on estimating the longitudinal
effect of initiating treatment for opioid use disorder (OUD) with extended-release naltrexone (XR-NTX) vs. buprenorphine-naloxone (BUP-NX) on risk of illicit opioid use during the fourth week of treatment 
that operates through the mediator of craving symptoms, using
symptoms of depression
and withdrawal as time-varying covariates. We include weekly measures for each of the first 4 weeks of treatment.

\section{Notation and definition of LMTP interventional (in)direct
  effects}\label{sec:notation}

Let $X_1,\ldots, X_n$ denote a sample of i.i.d. observations with
$X=(L_1, A_1, Z_1, M_1, L_2, \ldots, A_\tau, Z_\tau, \allowbreak
M_\tau, L_{\tau+1})\sim \P$, where $A_t$ denotes a vector of
intervention variables such as treatment and/or loss-to-follow-up,
$Z_t$ denotes intermediate confounders, $M_t$ denotes a mediator of
interest, and $L_t$ denotes time-varying covariates. The outcome of
interest is a variable $Y= L_{\tau+1}$ measured at the end of the
study. We let $\P f = \int f(x)\dd \P(x)$ for a given function
$f(x)$. We use $\Pn$ to denote the empirical distribution of
$X_1,\ldots\,X_n$, and assume $\P$ is an element of the nonparametric
statistical model defined as all continuous densities on $X$ with
respect to a dominating measure $\nu$. We let $\E$ denote the
expectation with respect to $\P$, i.e.,
$\E\{f(X)\} = \int f(x)\dd\P(x)$. We also let $||f||^2$ denote the
$L_2(\P)$ norm $\int f^2(x)\dd\P(x)$. We use
$\bar W_t = (W_1,\ldots, W_t)$ to denote the past history of a
variable $W$, use $\ubar W_t = (W_t,\ldots, W_\tau)$ to denote the
future of a variable, and use
$H_{A,t} = (\bar L_t, \bar M_{t-1}, \bar Z_{t-1}, \bar A_{t-1})$
denote the history of all variables up until just before $A_t$. The
random variables $H_{Z,t}$, $H_{M,t}$, and $H_{L,t}$ are defined
similarly as $H_{Z,t}=(A_t, H_{A,t})$, $H_{M,t}=(Z_t, H_{Z,t})$, and
$H_{L,t}=(M_{t-1}, H_{M,t-1})$. For the complete history and past of a
random variable, we sometimes simplify $\bar W_\tau$ and $\ubar W_1$
as $\bar W$. By convention, variables with an index $t\leq 0$ are
defined as null, expectations conditioning on a null set are marginal,
products of the type $\prod_{t=k}^{k-1}b_t$ and $\prod_{t=0}^0b_t$ are
equal to one, and sums of the type $\sum_{t=k}^{k-1}b_t$ and
$\sum_{t=0}^0b_t$ are equal to zero. We let
$\g_{A,t}(a_t \mid h_{A,t})$ denote the probability mass function of
$A_t$ conditional on $H_{A,t}=h_{A,t}$, and assume $A_t$ takes values
on a finite set. The function $\g_{M,t}(m_t \mid h_{M,t})$ is defined
similarly, and we assume also that $M_t$ takes values on a finite
set. The variables $L_t$ and $Z_t$ are allowed to take values on any
set; i.e., they can be multivariate, continuous,
etc.



\begin{figure}[!htb]
  \centering
  \begin{tikzpicture}
    \Vertex{-1, 0}{W}{$L_1$}
    \Vertex{1, 0}{A1}{$A_1$}
    \Vertex{2, 0}{M1}{$M_1$}
    \Vertex{3, 0}{Z1}{$L_2$}
    \Vertex{2, -1}{L1}{$Z_1$}
    \ArrowB{W}{A1}{black}
    \Arrow{A1}{M1}{black}
    \Arrow{A1}{L1}{black}
    \Arrow{M1}{Z1}{black}
    \Arrow{L1}{M1}{black}
    \Arrow{L1}{Z1}{black}
    \ArrowL{A1}{Z1}{black}

    \Vertex{5, 0}{A2}{$A_2$}
    \Vertex{6, 0}{M2}{$M_2$}
    \Vertex{7, 0}{Z2}{$L_3$}
    \Vertex{6, -1}{L2}{$Z_2$}
    \node (dots) at (8, 0) {$\cdots$};
    \Arrow{A2}{M2}{black}
    \Arrow{A2}{L2}{black}
    \Arrow{M2}{Z2}{black}
    \Arrow{L2}{M2}{black}
    \Arrow{L2}{Z2}{black}
    \ArrowL{A2}{Z2}{black}

    \ArrowB{Z1}{A2}{black}

    \Vertex{9, 0}{At}{$A_\tau$}
    \Vertex{10, 0}{Mt}{$M_\tau$}
    \Vertex{11, 0}{Zt}{$Y$}
    \Vertex{10, -1}{Lt}{$Z_\tau$}
    \Arrow{At}{Mt}{black}
    \Arrow{At}{Lt}{black}
    \Arrow{Mt}{Zt}{black}
    \Arrow{Lt}{Mt}{black}
    \Arrow{Lt}{Zt}{black}
    \ArrowL{At}{Zt}{black}
  \end{tikzpicture}
  \caption{Directed acyclic graph. For simplicity, the symbol
    $\boldsymbol{\mapsto}$ is used to indicate arrows from all nodes
    in its left to all nodes in its right.}
  \label{fig:dag}
\end{figure}
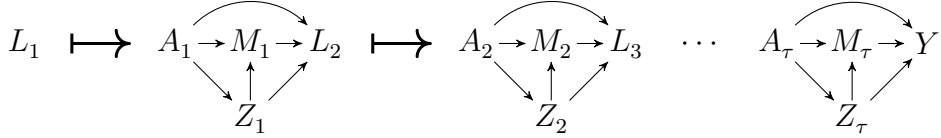

We formalize the definition of the causal effects using a
non-parametric structural equation model
\citep{Pearl00}. Specifically, for each time point $t$, we assume the
existence of deterministic functions $f_{A,t}$, $f_{Z,t}$, $f_{M,t}$,
and $f_{L,t}$ such that $A_t=f_{A,t}(H_{A, t}, U_{A,t})$,
$Z_t=f_{Z,t}(H_{Z,t}, U_{Z,t})$, $M_t=f_{M,t}(H_{M,t}, U_{M,t})$, and
$L_t=f_{L,t}(H_{L,t}, U_{L,t})$. Here
$U=(U_{A,t}, U_{Z,t}, U_{M,t}, U_{L,t}, U_Y:t\in \{1,\ldots,\tau\})$
is a vector of exogenous variables, with unrestricted joint
distribution. This model can be expressed in the form of a Directed
Acyclic Graph (DAG) as in Figure~\ref{fig:dag}.
In this paper, we are concerned with the definition and estimation of
the causal effect of an intervention on $\bar A$ on $Y$,
as well as its decomposition in terms of the effect through all paths
involving the mediator $\bar M$ versus effects through all other
mechanisms that do not involve any component of $\bar M$.  Mediation effects will be defined in terms of
hypothetical interventions where the equations
$A_t=f_{A,t}(H_{A,t}, U_{A,t})$ and $M_t=f_{M,t}(H_{M,t}, U_{M,t})$
are removed from the structural model, and the treatment and mediator
nodes are externally assigned as follows.  Let $\bar M(\bar a)$ denote the
counterfactual mediator vector observed in a hypothetical world where
$\bar A = \bar a$.  For a value $\bar m$ in the range of $\bar M$ we
also define a counterfactual outcome $Y(\bar a, \bar m)$ as the
outcome in a hypothetical world where $\bar A=\bar a$ and
$\bar M = \bar m$. These variables are defined in terms of our assumed
NPSEM in the supplementary materials.

Let $\bar a^\star=( a^\star_1,\ldots, a^\star_\tau)$ and
$\bar a'=( a'_1,\ldots, a'_\tau)$ denote two user-specified values in
the range of $\bar A$, and let $\bar G(\bar a)$ denote a random draw
from the distribution of the counterfactual variable $\bar M(\bar
a)$. We define the conditional interventional effect as
$\E[Y(\bar a', \bar G(\bar a')) - Y(\bar a^*, \bar G(\bar a^*))]$, and
decompose it into an interventional direct and indirect effect as follows:
\begin{align}
  \E&[Y(\bar a', \bar G(a')) - Y(\bar a^\star, \bar G(\bar a^\star))] =\notag\\
    & \underbrace{\E[Y(\bar a', \bar G(\bar a')) - Y(\bar a',
      \bar G(\bar a^\star))]}_{\text{Indirect effect (through $\bar M$)}} +
      \underbrace{\E[Y(\bar a', \bar G(\bar a^\star)) - Y(\bar a^\star, \bar G(\bar a^\star))]}_{\text{Direct
      effect (not through $\bar M$)}}.\label{eq:decomp}
\end{align}
This is the definition of interventional effect in a longitudinal
setting given by \cite{vanderweele2017mediation}. In what follows we
focus on identification and estimation of the parameters
$\E[Y(\bar a', G(\bar a^\star))]$ for fixed $\bar a'$, $\bar a^\star$,
from which we can get the effects in (\ref{eq:decomp}). Issues related
to the interpretation of these direct and indirect effects are
discussed at length elsewhere \citep[e.g.,][]{moreno2018understanding}.

The above setup allows for the definition of causal effects for
time-to-event outcomes subject to loss-to-follow-up or censoring as
follows. Let $A_t=(A_{1,t},A_{2,t})$, where $A_{1,t}$ denotes the
exposure at time $t$, $A_{2,t}$ is equal to one if the unit remains
uncensored at time $t+1$ and zero otherwise, and $Y=L_{\tau+1}$
denotes event-free status at the end of study follow-up. Assume
monotone loss-to-follow-up so that $A_{2,t}=0$ implies $A_{2,k}=0$ for
all $k>t$, in which case all the data for $k > t$ become
degenerate. In this case we could define the effects as above with
$\bar a'=((\bar a_{1,1}',1),\ldots,(\bar a_{1,\tau}',1))$ and
$\bar a^\star=((\bar a_{1,1}^\star,1),\ldots,(\bar
a_{1,\tau}^\star,1))$, contrasting regimes where treatment at time $t$
is set to $A_{1,t}=a_{1,t}'$ vs $A_{1,t}=a_{1,t}^\star$ while setting
censoring status $A_{2,t}=1$ as not censored for everyone. This
definition of causal effect for time-to-event outcomes in terms of
interventional effects bypasses some (but not all) problems that occur
with natural effects due to the fact that a counterfactual
longitudinal mediator may be truncated by death, which may render the
counterfactual survival time undefined \citep{lin2017mediation,
  huang2017causal,didelez2019defining}.

In what follows we will use the notation $H_{A,t}'$, $H_{Z,t}'$,
$H_{M,t}'$, and $H_{L,t}'$ to refer to the intervened histories of the
variables. For example,
$H_{L,t}' =(\bar L_{t-1}, \bar M_{t-1}, \bar Z_{t-1}, \bar
A_{t-1}=\bar a_{t-1}')$. The histories $H_{A,t}^\star$,
$H_{Z,t}^\star$, $H_{M,t}^\star$, and $H_{L,t}^\star$ are defined
analogously. The following assumptions will be sufficient to prove
identification of the parameter $\E[Y(\bar a', G(\bar a^\star))]$:

\begin{assumptioniden}[Conditional exchangeability of treatment and
  mediator assignment]\label{ass:iden} Assume:
  \begin{enumerate}[label=(\roman*)]
  \item $A_t\indep Y(\bar a', \bar m)\mid H_{A,t}'$ for all $t$, $\bar m$. \label{ass:ceAY}
  \item $M_t\indep Y(\bar a', \bar m)\mid H_{M,t}'$ for all $t$, $\bar m$. \label{ass:ceMY}
  \item $A_t\indep \ubar M_t(\bar a^\star)\mid H_{A,t}^\star$ for all $t$. \label{ass:ceAM}
  \end{enumerate}
\end{assumptioniden}

\begin{assumptioniden}[Positivity of treatment and mediator assignment
  mechanisms]\label{ass:pos} Assume:
  \begin{enumerate}[label=(\roman*)]
  \item $\P\{\g_{A,t}(a_t'\mid H_{A,t}')>0\}=1$ and
    $\P\{\g_{A,t}(a_t^\star\mid H_{A,t}^\star)>0\}=1$ for all
    $t$, \label{ass:posA}
  \item If $\P\{\g_{M,t}(m_t\mid H_{M,t}^\star) > 0\}=1$ then
    $\P\{\g_{M,t}(m_t\mid H_{M,t}') > 0\}=1$ for all $t$ and
    $m_t$. \label{ass:posM}
  \end{enumerate}
\end{assumptioniden}
Assumptions \ref{ass:iden}\ref{ass:ceAY},
\ref{ass:iden}\ref{ass:ceMY}, and \ref{ass:iden}\ref{ass:ceAM}
together with the assumed DAG in Figure~\ref{fig:dag} state that
$H_{A,t}$ contains all the common causes of $A_t$ and $Y$, $H_{M,t}$
contains all the common causes of $M_t$ and $Y$, and $H_{A,t}$
contains all the common causes of $A_t$ and $M_t$, respectively. These
are the standard assumptions of no-unmeasured confounders for the
treatment-outcome, mediator-outcome, and treatment-mediator
relations. Assumption~\ref{ass:pos}\ref{ass:posA} states that there is
enough randomness in the treatment assignment processes at each time
point so that both treatment regimes $a'_t$ and $a^\star_t$ can occur
for all covariate strata defined by
$H_{A,t}$. Assumption~\ref{ass:pos}\ref{ass:posM} states that if a
value of the mediator can occur in the group with $A_t=a^\star_t$,
then that mediator value can also occur in the group with $A_t=a'_t$,
conditional on other covariates in $H_{M,t}$. This assumption
precludes treatment and mediator assignments that occur
deterministically within strata of covariates. We have the following
identification result.
\begin{theorem}[Identification]\label{theo:iden}
  Under Assumptions~\ref{ass:iden} and~\ref{ass:pos}, the
  interventional effect $\theta=\E[Y(\bar a', G(\bar a^\star))]$ is
  identified as follows. Let
  \begin{align*}
    \varphi(\bar m) &= \int l_{\tau+1}
    \prod_{t=1}^{\tau}\p[l_{t+1}\mid h_{L,t+1}']\p[z_t\mid
                      h_{Z,t}']\p(l_1)\dd\nu(\bar l,\bar z)\\
    \lambda(\bar m)&=\int \prod_{t=1}^\tau\p[l_{t+1},m_t,
    z_t\mid h_{Z,t}^\star]\p(l_1)\dd\nu(\bar
                        l, \bar  z)  \end{align*}
  Then $\theta$ is identified as
  \begin{equation}
    \theta = \sum_{\bar m} \varphi(\bar m)\lambda(\bar m)\label{eq:iden}
  \end{equation}
\end{theorem}

The above identification result generalizes several identification
results in the literature. When $\tau=1$ and $Z=\emptyset$, this
identification formula is equal to the identification formula for the
natural direct and indirect effects as derived by \cite{Pearl01} under
an additional cross-world counterfactual assumption.  When $\tau=1$
and the confounder $Z$ is present, this identification formula is
equal to the identification formula for the interventional effect
described by \cite{vanderweele2014effect}. If $\tau >1$ and
$Z_t=\emptyset$ for all $t$, Equation~\ref{eq:iden} reduces to formula
(1) in \cite{vanderweele2017mediation}. If $\tau >1$ and
$L_t=\emptyset$ for $t\leq \tau$, Equation~\ref{eq:iden} reduces to
the identification result for Figure 5 given in page 926 of
\cite{vanderweele2017mediation}.

The identification formula in Equation~\ref{eq:iden} involves several
densities on $L_t$ and $Z_t$, which might be hard to estimate if these
variables take values on large sets (e.g., if they are continuous or
multivariate with a large dimension). To aid in estimation, we will
now discuss an alternative representation of Equation~\ref{eq:iden} in
the form of sequential regressions, which will allow us to construct
estimators for $\theta$ based on standard regression procedures. This
approach was first proposed by \cite{Bang05} and has become
standard in estimation of causal effects in longitudinal
studies \citep{luedtke2017sequential,rotnitzky2017multiply}.

Set $\Q_{Z, \tau+1} = Y$. For fixed values $\bar a'$ and $\bar m$, and
for $t=\tau, \ldots, 0$, recursively define the random variables
\begin{align}
  \Q_{L,t}(\bar H_{M,t},\ubar m_t)&= \E[\Q_{Z, t+1}(H_{A, t+1}, \ubar m_{t+1})\mid M_{t} = m_t, H_{M,t}]\label{eq:qM}\\
  \Q_{Z,t}(\bar H_{A,t},\ubar m_t) &= \E[\Q_{L,t}(\bar H_{M,t},\ubar m_t)\mid A_t =
             a_t' ,H_{A,t}],\label{eq:qZ}
\end{align}
To simplify notation, we will sometimes omit the dependence of the
above functions on $H_{A,t}$, $H_{M,t}$, and $\bar m$. In the proof of
Proposition \ref{prop:rep} (available in the Supplementary Materials)
we show that $\varphi(\bar m) = \Q_{L,0}(\bar m)$. The
counterfactual distribution $\lambda(\bar m)$ may be identified as
follows. Let $\Q_{M,\tau + 1} = 1$. For
$t=\tau, \ldots, 0$, recursively define
\[  \Q_{M, t}(\bar H_{A,t}, \ubar m_t)
  =\E[\one\{M_t =
    m_t\}\Q_{M, t+1}(\bar H_{A,t+1}, \ubar m_{t+1})\mid A_t=a_t^\star,
    H_{A,t}]\]
 Then, we have
$\lambda(\bar m) = \Q_{M, 0}(\bar m)$. This leads to the
following alternative expression for the mediational g-computation
formula in terms of sequential regressions:
\begin{prop}[Sequential regression representation of the longitudinal mediation g-formula]\label{prop:rep}
For $\theta$ defined in Theorem~\ref{theo:iden}, we have  
\[\theta = \sum_{\bar m} \Q_{L,0}(\bar m)\Q_{M, 0}(\bar
  m).\]
\end{prop}

This alternative expression of the longitudinal mediation formula
allows the construction of an estimator by sequential regression,
estimating the parameters $\Q_{L, t}(\bar m)$ and $\Q_{Z,t}(\bar m)$
sequentially for $t=\tau,\ldots,1$ to obtain an estimate of
$\Q_{L,0}(\bar m)$, and then estimating $\Q_{M, t}(\bar m)$
sequentially for $t=\tau,\ldots,1$ to obtain an estimate of
$\Q_{M,0}(\bar m)$.

In this paper we will estimate several sequential regressions similar
to those in (\ref{eq:qM}) and (\ref{eq:qZ}). There are at least two
alternatives to estimate these sequential regressions. The first is to
perform sequential regression separately for each $\bar m$ in the
range of $\bar M$. The second is to construct a pooled dataset where
we pool all values $\bar m$ in the range of $\bar M$ to obtain the
estimates for all $\bar m$ from a single sequential regression
procedure. In this paper we pursue the second approach (see
Algorithm~\ref{algo:cap}).

If the pooled sequential regressions are performed within a-priori
correctly specified parametric models, then estimators of $\theta$
based on Proposition~\ref{prop:rep} may be shown to be CAN, and the
Delta method or the non-parametric bootstrap may be used to construct
confidence intervals. However, positing correct parametric models for
the sequential regressions involved is generally unattainable a-priori
and requires model selection techniques even when parametric models
offer an appropriate fit. Furthermore, in most cases, especially with
a large number of variables, data-adaptive regression (e.g., machine
learning) tools that offer flexibility in modeling non-linearities and
interactions are necessary to attain consistency of the sequential
regressions and therefore consistency of the estimator of $\theta$.

Under model selection or data-adaptive regression, the sampling
distribution of the above sequential regression estimator is generally
unknown, which hinders computation of confidence intervals and other
uncertainty measures.

In the next section, we discuss efficiency theory for estimation of
$\theta$, which will allow the use of data-adaptive regression
techniques while also allowing the computation of valid (under
assumptions) standard errors and confidence intervals. The foundations
of our estimation approach are in semi-parametric efficient estimation
theory \citep[e.g.,][]{mises1947asymptotic, vanderVaart98,
  robins2009quadratic, Bickel97} and in the theory for doubly robust
estimation of causal effects using sequential regression
\citep[e.g.,][]{Robins00,Robins&Rotnitzky&Zhao94,vanderLaan2003,Bang05,
  vdl2006targeted, vanderLaanRose11, vanderLaanRose18,
  luedtke2017sequential,rotnitzky2017multiply}. Central to this theory
is the study of the non-parametric efficient influence function (EIF)
or canonical gradient, which characterizes the efficiency bound of the
longitudinal mediation functional given in Equation (\ref{eq:iden})
and allows the development of estimators under slow convergence rates
for the nuisance parameters involved
\citep{robins2009quadratic}. Specifically, our estimators will involve
finding von-Mises-type approximations for the parameters
$\Q_{L, t}(\bar m)$, $\Q_{Z, t}(\bar m)$, and $\Q_{M, t}(\bar m)$,
which can be intuitively understood as first order expansions with
second-order error remainder terms. Because the errors in the
expansion are second-order, this will mean that the resulting
estimator of $\hat\theta$ will be consistent and asymptotically normal
at rate $n^{1/2}$ as long as the second-order error terms converge to
zero at rate $n^{1/2}$. This convergence rate would be satisfied, for
example, if the all regression functions used for estimation converge
at rate $n^{1/4}$. We will elaborate on this discussion in
\sec\ref{sec:estima} when we present the asymptotic normality theorems
for the proposed estimators.

\section{Efficiency theory}\label{sec:eff}
Define the following random variables
\[\G_{A,t}(H_{A,t})'=\frac{\one\{A_t=a_t'\}}{\g_t(a_t' \mid
    H_{A,t})},\,\,\G_{A,t}^\star(H_{A,t})=\frac{\one\{A_t=a_t^\star\}}{\g_t(a_t^\star\mid
    H_{A,t})},\,\,\,
  \G_{M,t}(H_{M,t},m_t)=\frac{\one\{M_t=m_t\}}{\g_{M,t}(M_t\mid
    H_{M,t})},\] where, as before, we will sometimes omit the
dependence on $H_{A,t}$, $H_{M,t}$, and $m_t$. For $t=0,\ldots,\tau$
and for $k=1,\ldots,\tau$, and for a nuisance parameter 
$\eta = \{\G_{A,t}',\G_{A,t}^\star,\G_{M,t}, \Q_{Z,t}, \Q_{M,t},
\Q_{L,t}:t=1,\ldots,\tau\}$, define 
\[\C_{l,u}' =
  \prod_{r=l}^{u}\G_{A,r}',\quad \C_{l,u}^\star = 
  \prod_{r=l}^{u}\G_{A,r}^\star,\quad \H_{l,u} = 
  \prod_{r=l}^{u}\G_{M,r}\]
and {\footnotesize
  \begin{align}
    \D_{L,t}(\ubar X_t, \ubar m_t)
    & =\sum_{s=t}^\tau
      \C_{t+1,s}'\H_{t,s}
      \left\{\Q_{Z,s+1} -
      \Q_{L,s}\right\}
      +\sum_{s=t+1}^\tau \C_{t+1,s}'\H_{t,s-1}\left\{\Q_{L,s} -
      \Q_{Z,s}\right\} + \Q_{L,t}.\label{eq:Dl}\\
    \D_{Z,t}(\ubar X_t, \ubar m_t)
    & =\sum_{s=t}^\tau \C_{t,s}'\H_{t,s}   
      \left\{\Q_{Z,s+1} -
      \Q_{L,s}\right\}
      +\sum_{s=t}^\tau \C_{t,s}'\H_{t,s-1}\left\{\Q_{L,s} -
      \Q_{Z,s}\right\}  + \Q_{Z,t}\label{eq:Dz}\\
    \D_{M,t}(\ubar X_t, \ubar m_t)
    &= \sum_{s=t}^\tau\C_{t,s}^\star\left(\prod_{k=t}^{s-1}\one(M_k=
      m_k)\right)
      \left\{\one(M_s=
      m_s)\Q_{M,s+1} - \Q_{M,s}\right\} + \Q_{M,t}.\label{eq:Dm}    
  \end{align}}%
Whenever necessary, we make explicit the
dependence of these functions on $\eta$ using notation such as
$\D_{Z,t}(\eta)$ or $\D_{Z,t}(\ubar X_t, \ubar m_t;\eta)$.

\begin{lemma}[von-Mises-type approximation for $\Q_{Z,t}$,
  $\Q_{L,t}$, and $\Q_{M,t}$]\label{lemma:vm}
  Let $\tilde\eta$ denote an arbitrary value of $\eta$. For second
  order terms $\Rem_{L, t}(\eta,\tilde \eta)$,
  $\Rem_{Z, t}(\eta,\tilde \eta)$, and $\Rem_{M,t}(\eta,\tilde \eta)$
  we have the following first order expansions:
  \begin{align*}
    \Q_{L,t}
    &=\E[\D_{Z,t+1}(\tilde \eta)
      \mid  M_t = m_t, H_{M,t}] +
      \Rem_{L, t}(\eta,\tilde \eta)\\
    \Q_{Z,t}
    &=\E[\D_{L,t}(\tilde \eta)\mid  A_t=a_t', H_{A,t}] +
      \Rem_{Z, t}(\eta,\tilde \eta),
  \end{align*}
where we let $\D_{Z,\tau+1}(\tilde \eta) = Y$. For $H_{A,t}$ such that
$\bar M_{t-1}=\bar m_{t-1}$, we also have
\[ \Q_{M,t} =\E[\one\{M_t=m_t\}\D_{M,t+1}(\tilde \eta) \mid A_t=a_t^\star, H_{A,t}] +
  \Rem_{M,t}(\eta,\tilde \eta),
\]
where we let $\D_{M,\tau+1}(\tilde \eta)=1$.

\end{lemma}

The terms $\Rem_{L, t}(\eta,\tilde \eta)$,
$\Rem_{Z, t}(\eta,\tilde \eta)$, and $\Rem_{M,t}(\eta,\tilde \eta)$
are second-order error terms involving expectations of products of
errors such as
$(\tilde\G_{M,s} - \G_{M,s})(\tilde \Q_{L,s}-\Q_{L,s})$,
$(\tilde\G_{A,s}'- \G_{A,s}')(\tilde \Q_{Z,s}-\Q_{Z,s})$, and
$(\tilde\G_{A,s}^\star - \G_{A,s}^\star)(\tilde \Q_{M,s}-\Q_{M,s})$,
and their explicit form is given in the supplementary materials. This
lemma and specifically the second-order form of these remainder terms
has important implications in terms of estimation. Specifically, this
lemma says that for any value $\tilde\eta$, which could represent an
inconsistent estimator, regressing $\D_{Z,t+1}(\tilde \eta)$ on
$H_{M,t}$ among units with $M_t=m_t$ yields a consistent estimator of
$\Q_{L,t}$, as long as $\Rem_{L, t}(\eta,\tilde \eta)$ is
small. Because $\Rem_{L, t}(\eta,\tilde \eta)$ is a sum of products of
errors, it may be reasonable to assume that this term is small for
data-adaptive regression estimators of $\eta$. Specifically, in
\sec\ref{sec:estima}, consistency and asymptotic normality of the
estimators will require that $\Rem_{L, t}(\eta,\hat \eta)$ converges
to zero in probability at rate $n^{-1/2}$ or faster. The second-order
term structure of this remainder term means that this fast convergence
rate can be achieved under slower convergence rates for each of the
components of $\hat\eta$. For example, it will be achievable if all
the components of $\hat\eta$ converge in probability to their correct
limits at rate $n^{-1/4}$. This kind of rate is achievable by several
data-adaptive regression methods, such $\ell_1$ regularization
\citep{bickel2009simultaneous}, regression trees
\citep{wager2015adaptive}, neural networks \citep{chen1999improved},
or the highly adaptive lasso \citep{benkeser2016highly}. Analogous
considerations apply to estimation of $\Q_{Z,t}$ and $\Q_{M,t}$.

In addition, an application of the Delta method along with
Lemma~\ref{lemma:vm} yields the following efficient influence function
(EIF) for estimation of $\theta$ in the non-parametric model (see the
supplementary materials for a proof):

\begin{theorem}[Efficient influence function for
  $\theta$]\label{theo:eif}
  The EIF for $\theta$ in the non-parametric model is given by
  \[\S(X,\eta)= \sum_{\bar m\in \bar{\mathcal M}}\left[\{\D_{Z,1}(X,\bar m;\eta)
    - \varphi(\bar m)\}\lambda(\bar m) + \{\D_{M,1}(X,\bar m;\eta)
    - \lambda(\bar m)\}\varphi(\bar m)\right].\]
\end{theorem}

This implies that the non-parametric efficiency bound for estimation
of $\theta$ is $\var[\S(X,\eta)]$, and that an efficient estimator of
$\theta$ will satisfy
\[\sqrt{n}(\hat\theta-\theta)=\frac{1}{\sqrt{n}}\sum_{i=1}^n\S(X_i,\eta)
  +o_\P(1),\]
licensing the construction of Wald-type confidence intervals and
hypothesis tests based on the central limit theorem. In the following
section we will describe an algorithm to construct such an estimator. 
\section{Efficient estimation using sequential doubly robust
  regression}\label{sec:estima}

For any natural number $n$ we denote
$[n]=\{1,\ldots,n\}$. Furthermore, let $\mathcal D=\{X_i: i\in [n]\}$
denote the observed dataset, and let
$\mathcal M_t=\{m_j: j\in [J_t]\}$ denote the set of unique values
that the mediator $M_t$ takes in the sample $\mathcal D$ at time point
$t$, where $J_t\leq n$ is the number of unique values. 

Estimating
$\Q_{L,0}(\bar m)$ for each $\bar m$ would in principle require
fitting $\prod_{k=1}^\tau J_k$ sets of $\tau$ sequential regressions
as in formula (\ref{eq:qM}), one set of sequential regressions for
each one of the possible values of $\bar m$. This can be
computationally prohibitive. For example, estimating
$\Q_{L,0}(\bar m)$ in a study with $\tau=5$ and a mediator taking on
three different values involves $3^5=243$ sets of five sequential
regressions, each in a dataset of size $n$. To alleviate computational
complexity, we propose to fit regressions in pooled datasets
constructed sequentially from $t=\tau$ to $t=1$, where at time point $t$, each
observation in the dataset at time point $t+1$ is repeated $J_t$
times. Specifically, consider the original dataset
$\mathcal D=\{X_1,\ldots,X_n\}$. Regressions at time point $\tau$ are
constructed in an augmented dataset defined as
{\small\[\mathcal D_\tau^+ = \mathcal D\times \mathcal
    M_\tau=\{(X_1,m_1), \ldots, (X_1,m_{J_\tau}), \ldots,(X_\tau,m_1),
    \ldots, (X_\tau,m_{J_\tau})\},\]}%
where $\times$ denotes Cartesian product. Regressions at time point
$\tau-1$ are constructed in a dataset defined as
$\mathcal D_{\tau-1}^+=\mathcal D_\tau^+\times \mathcal M_{\tau-1}$, and
so on. Thus, an observation in dataset ${\cal D}_t^+$ is a duple
$(X_i,\ubar m_t)$. 

Under this data pooling approach estimating $\Q_{L,0}$ in a study with
$\tau=5$ and a mediator taking on three different values involves five
sequential regressions, each in datasets of size $3n$, $3^2n$, $3^3n$,
$3^4n$, $3^5n$, respectively, where the predictor set decreases in
size as the dataset increases in size. The specific details of this
sequential regression algorithm using pooled datasets are described in
Algorithm~\ref{algo:cap}. An R package implementing the algorithm is
available at \url{https://github.com/nt-williams/lcm}. In brief, the
algorithm proceeds sequentially from $t=\tau,\ldots,1$ performing
regression of pseudo-outcomes given by $\D_{Z,t+1}(\hat\eta)$,
$\D_{L,t}(\hat\eta)$, and $\one\{M_t=m_t\}\D_{M,t+1}(\eta)$, where
these regressions are cross-fitted and based on datasets
$\mathcal D_t^+$ that pool over values of $m_t$ as defined above.

In addition to using the above pooled datasets to construct sequential
regressions, we will use cross-fitting, which helps to avoid imposing
entropy conditions on the initial estimators
\citep{bickel1982adaptive,klaassen1987consistent,zheng2011cross,
  chernozhukov2018double}, while yielding estimators that are
CAN. This allows us to use flexible regression algorithms from the
statistical and machine learning literature, which may be better at
capturing the true functional form of the regression functions and
thus making the error terms $\Rem_{L,0}(\hat\eta,\eta)$,
$\Rem_{Z,0}(\hat\eta,\eta)$, and $\Rem_{M,0}(\hat\eta,\eta)$
small. Let ${\cal P}_1, \ldots, {\cal P}_V$ denote a random partition
of the data set $\mathcal D$ into $V$ prediction sets of approximately
the same size. That is, ${\cal P}_v\subset \{1, \ldots, n\}$;
$\bigcup_{j=1}^J {\cal P}_v = {\cal D}$; and
${\cal P}_v\cap {\cal P}_{v'} = \emptyset$. In addition, for each $v$,
the associated training sample is given by
${\cal T}_v = {\cal D} \setminus {\cal P}_v$. The cross-fitting
algorithm is described in Algorithm~\ref{algo:cf}.

Our proposed estimation algorithm 
 satisfies sequential double
robustness in the sense of Lemma~\ref{lemma:dr} below \citep{luedtke2017sequential,rotnitzky2017multiply,
  diaz2021nonparametricmtp}. To illustrate sequential double robustness,
consider an alternative estimation strategy in which the parameters
$\Q_{L,t}$, $\Q_{Z,t}$, and $\Q_{M,t}$ are estimated directly (i.e.,
based on (\ref{eq:qM}) and (\ref{eq:qZ})) using flexible regression techniques,
and the estimators of $\varphi(\bar m)$ and $\lambda(\bar m)$ are
constructed by taking the empirical mean of
$\D_{L,1}(X,\bar m;\hat\eta)$ and $\D_{M,1}(X,\bar m;\hat\eta)$. Note,
however, that the functions $\Q_{L,t}$, $\Q_{Z,t}$, and $\Q_{M,t}$ are
functions of the sequences $\{\Q_{L,t+1},\ldots,\Q_{L,\tau}\}$,
$\{\Q_{Z,t+1},\ldots,\Q_{Z,\tau}\}$, and
$\{\Q_{M,t+1},\ldots,\Q_{M,\tau}\}$. It therefore appears that
consistent estimation of these parameters at time $t$ requires
consistent estimation of these future sequences.

Sequential doubly robust estimators decouple estimation of conditional
expectations at time $t$ from consistent estimation of sequences of
estimators at time $t+1,\ldots,\tau$, therefore achieving extra
robustness. To introduce sequential double robustness, define the data
dependent parameters
\begin{align*}
  \check\Q_{L,t} &= \E[\D_{Z,t+1}(\hat\eta)\mid M_t=m_t, H_{M,t}]\\
  \check\Q_{Z,t} &= \E[\D_{L,t}(\hat\eta)\mid A_t=a_t', H_{A,t}]\\
  \check\Q_{M,t} &= \E[\one\{M_t=m_t\}\D_{M,t+1}(\hat\eta)\mid A_t=a_t^\star, H_{A,t}],
\end{align*}
where the expectation is with respect to the distribution of $X$,
i.e., the estimator $\hat\eta$ is considered fixed. Then we have the
following result:

\begin{lemma}[Sequential double robustness]\label{lemma:dr}
  Assume that, at each time point $t$, we have
  \begin{enumerate}[label=(\roman*)]
  \item $||\hat
    \G_{A,t}'-\G_{A,t}'||=o_P(1)$ or $||\hat
    \Q_{Z,t}-\check\Q_{Z,t}||=o_P(1)$, and
  \item $||\hat
    \G_{M,t}-\G_{M,t}||=o_P(1)$ or $||\hat
    \Q_{L,t}-\check\Q_{L,t}||=o_P(1)$, and
  \item $||\hat
    \G_{A,t}^\star-\G_{A,t}^\star||=o_P(1)$ or $||\hat
    \Q_{M,t}-\check\Q_{M,t}||=o_P(1)$,
  \end{enumerate}
  then we have $\hat\theta = \theta + o_P(1)$, for $\hat\theta$
  defined in Algorithm~\ref{algo:cap}.
\end{lemma}
Here we note that the error terms $||\hat \Q_{Z,t}-\check\Q_{Z,t}||$,
$||\hat \Q_{L,t}-\check\Q_{L,t}||$, and
$||\hat \Q_{M,t}-\check\Q_{M,t}||$ only depend on the consistency of
the regression procedure used at time point $t$ and not on the
consistency of estimators at any time point $s>t$. This lemma is the
result of an application of Lemma 4 in
\cite{diaz2021nonparametricmtp}. Furthermore, the sequential
regression estimator $\hat\theta$ satisfies the following weak
convergence result:
\begin{theorem}[Weak convergence of $\hat\theta$]\label{theo:weak}
  Assume that, at each time point $t$, we have
  \begin{enumerate}[label=(\roman*)]
  \item $||\hat
    \G_{A,t}'-\G_{A,t}'||\,||\hat
    \Q_{Z,t}-\check\Q_{Z,t}||=o_P(n^{-1/2})$, and
  \item $||\hat
    \G_{M,t}-\G_{M,t}||\,||\hat
    \Q_{L,t}-\check\Q_{L,t}||=o_P(n^{-1/2})$, and
  \item $||\hat
    \G_{A,t}^\star-\G_{A,t}^\star||\,||\hat
    \Q_{M,t}-\check\Q_{M,t}||=o_P(n^{-1/2})$, and
  \item $\P(\G_{A,t}' < c) = \P(\G_{A,t}^\star < c) = \P(\G_{M,t} < c) =
    1$ for some $c<\infty$. 
  \end{enumerate}
  Then we have \[\sqrt{n}(\hat\theta-\theta) \rightsquigarrow N(0, \sigma^2),\]
  where $\sigma^2=\var[\S(X;\eta)]$ is the non-parametric efficiency bound.
\end{theorem}
Note that the weak convergence of $\hat\theta$ requires consistent
estimation of the sequential regression functions at the rates stated
in the theorem. Data-adaptive regression methods avoid reliance on
parametric assumptions to achieve consistent estimation.

Theorem~\ref{theo:weak} allows the construction of confidence
intervals as $\hat\theta \pm z_{\alpha/2}\hat\sigma/\sqrt{n}$, where
$\hat\sigma^2$ is the empirical variance of $\S(X;\hat\eta)$ and
$z_\alpha$ is the quantile of a standard normal distribution.

\SetKwFunction{Extract}{Extract}
\SetKwFunction{CrossFit}{CrossFit} \SetKwFunction{Regress}{Regress}
\SetKwFunction{Predict}{Predict} \SetKwFunction{Append}{AppendColumns}
\SetKwFunction{Subset}{Subset} \SetKwFunction{Mean}{Mean}
\SetKwFunction{Var}{Variance} \SetKw{KwRet}{Return}
\SetKwInput{KwInput}{Input} \SetKwProg{Fn}{Function}{:}{}
\SetKwData{fit}{fit}\SetKwData{out}{out}
\begin{algorithm}[!htb]
  \caption{Cross-fitted regression}\label{algo:cf}
  \KwInput{$\cal T$: a training dataset; $\cal P$: a prediction
    dataset; $Y$: an outcome; $X$ a set of predictors; type: a type of
  prediction (e.g., probability or mean outcome (the default))}
  \Fn{\CrossFit{$\cal T$, $\cal P$, $Y$, $X$, type}}{
    $\fit\gets \Regress(\text{outcome} = Y, \text{predictors} = X,
    \text{training data} = {\cal T})$\;
    $\out\gets \Predict(\fit, \text{data} = {X\in \cal P}, \text{type} = \text{type})$\;
  }
  \KwRet $\out$
\end{algorithm}
  
\begin{algorithm}[!htb]
  \caption{Cross-fitted efficient estimation of
    $\theta$}\label{algo:cap}
  {\small
  Split the data set $\mathcal D$  randomly into $V$ parts $\mathcal P_v$ of
  approximately the same size\;
  $\mathcal P_{v,\tau+1}^+\gets  \mathcal P_v$ \text{for all $v$}\;
  $\hat \D_{Z,\tau+1}\gets Y$\;
  $\hat \D_{M,\tau+1}\gets 1$\;
  \For{$t = \tau,\ldots,1$}
  {
    \For{$v\in [V]$}
    {
      $\mathcal T_{v,t}^+\gets  \bigcup_{v'\neq v} \mathcal
      P_{v',t+1}^+ \times \mathcal M_t$\;
      $\mathcal P_{v,t}^+\gets  \mathcal P_{v,t+1}\times \mathcal
      M_t$\;

      $\hat\Q_{L,t}\gets \CrossFit(\Subset(\mathcal T_{v,t}^+,
      M_t=m_t), \mathcal P_{v,t}^+,\hat \D_{Z,t+1},(\ubar m_t,
      H_{M,t}))$\;
      $\hat\g_t\gets \CrossFit(\mathcal T_v, \mathcal P_{v,t}^+,
      A_t, H_{A,t},\text{probability})$\;
      $\hat\g_{M,t}\gets \CrossFit(\mathcal T_v, \mathcal P_{v,t}^+,
      M_t, H_{M,t},\text{probability})$\;
      $\hat\G_{A,t}'\gets
      \one\{A_t=a_t'\}/\hat \g_t(A_t\mid H_{A,t})$\;
      $\hat\G_{A,t}^\star\gets
      \one\{A_t=a_t^\star\}/\hat \g_t(A_t\mid H_{A,t})$\;
      $\hat\G_{M,t}\gets \one\{M_t=m_t\}/\hat\g_{M,t}(M_t\mid H_{M,t})$\;
      $\hat\D_{L,t} \gets \D_{L,t}(\hat\eta)$ using formula
      (\ref{eq:Dl}) and data $\mathcal P_{v,t}^+$\;
      $\mathcal P_{v,t}^+\gets \Append(\mathcal P_{v,t}^+,
      \hat\Q_{L,t}, \hat\G_{A,t}', \hat\G_{A,t}^\star, \hat\G_{M,t}, \hat\D_{L,t})$\;
    }
    \For{$v\in [V]$}
    {
      $\mathcal T_{v,t}^+\gets  \bigcup_{v'\neq v} \mathcal
      P_{v',t}^+$\;
      $\hat\Q_{Z,t}\gets \CrossFit(\Subset(\mathcal T_{v,t}^+,
      A_t=a_t'), \mathcal P_{v,t}^+,\hat \D_{L,t},(\ubar m_t,
      H_{A,t}))$\;
      $\hat\Q_{M,t}\gets \CrossFit(\Subset(\mathcal T_{v,t}^+,
      A_t=a_t^\star), \mathcal P_{v,t}^+,\one\{M_t=m_t\}\hat
      \D_{M,t+1},(\ubar m_t, H_{A,t}))$\;

      $\hat\D_{Z,t} \gets \D_{Z,t}(\hat\eta)$ using formula
      (\ref{eq:Dz}) and data $\mathcal P_{v,t}^+$\;
      $\hat\D_{M,t} \gets \D_{M,t}(\hat\eta)$ using formula
      (\ref{eq:Dm}) and data $\mathcal P_{v,t}^+$\;
      $\mathcal P_{v,t}^+\gets \Append(\mathcal P_{v,t}^+,
      \hat\Q_{Z,t}, \hat\Q_{M,t}, \hat\D_{Z,t},\hat\D_{M,t})$\;
    }
  }  
  \For{$\bar m^\star\in\bar{\mathcal M}$}
  {
    \For{$v\in [V]$}
    {
      $\IF_{\varphi,v}(\bar m^\star)\gets \Extract(\hat \D_{Z,1},
      \Subset(\mathcal P_{v,1}^+, \bar m=\bar m^\star))$\;
      $\IF_{\lambda,v}(\bar m^\star)\gets \Extract(\hat \D_{M,1}, \Subset(\mathcal P_{v,1}^+, \bar m=\bar m^\star))$\;
      $\hat\varphi_v(\bar m^\star)\gets \Mean(\hat \IF_{\varphi,v}(\bar m^\star))$\;
      $\hat\lambda_v(\bar m^\star)\gets \Mean(\hat \IF_{\lambda,v}(\bar m^\star))$\;
    }
    $\hat\varphi(\bar m)\gets 1/V\sum_v\hat\varphi_v(\bar m)$\;
    $\hat\lambda(\bar m)\gets 1/V\sum_v \hat\lambda_v(\bar m)$\;
  }
  $\hat\theta\gets \sum_{\bar m}\hat\varphi(\bar m)\hat\lambda(\bar
  m)$\;
  $\hat \S\gets \sum_{\bar m}\{(\IF_{\varphi}(\bar m) - \hat\varphi(\bar
  m))\hat\lambda(\bar m) + (\IF_{\lambda}(\bar m) - \hat\lambda(\bar
  m))\hat\varphi(\bar m)\}$\;
  $\hat\sigma^2\gets \Var(\hat \S)/n$
}%
\end{algorithm}



\section{Illustrative application}\label{sec:aplica}
We applied our proposed estimators to a longitudinal mediation
question from a comparative effectiveness trial of extended-release
naltrexone (XR-NTX) vs. buprenorphine-naloxone (BUP-NX) for the
treatment of opioid use disorder (OUD) \citep{lee2018comparative}.
Specifically, we were interested in estimating the extent to which
differences in use of illicit opioids during the first month of
treatment between the two medications was due to mediation by
self-reported craving of opioids, among those completing the
detoxification requirement and initiating treatment. This involved
estimating the interventional direct effect of being treated with
XR-NTX vs. BUP-NX on risk of using illicit opioids during the first
four weeks of treatment, not operating through differences in craving
of opioids; and the interventional indirect effect of being treated
with XR-NTX vs. BUP-NX on risk of using illicit opioids during the
first four weeks of treatment that did operate through differences in
craving.
 
Patients report less opioid use when on XR-NTX vs. BUP-NX (as well as
methadone) \citep{greiner2021naturalistic,solli2018effectiveness}, but
the reasons underlying this difference are not well understood. In
this study, patients were randomized to receive XR-NTX or BUP-NX. At time of randomization, a large number (over 30) baseline covariates, denoted $L_1$, (listed in the supplemental
materials) were measured. Although patient assignment to XR-NTX vs. BUP-NX was
randomized, we are estimating effects only among those who initiated
treatment. Treatment initiation is not randomized, and likely depends on
patient characteristics. Initiation is also more difficult for those assigned to XR-NTX, because it requires complete detoxification \citep{lee2018comparative}. Our adjustment for an extensive
set of possibly confounding variables ($L_1$) helps address lack of
randomization in the exposure groups. We use $A_1=1$ to denote initiation with XR-NTX and
$A_1=0$ to denote initiation with BUP-NX. The outcome of this study is opioid use as detected by weekly urine drug screen or Timeline Followback interview \citep{sobell1995alcohol}. 
We use $L_t$ to denote
opioid use measured at week $t+1$ for $t \in \{2, 3, 4\}$, which detects use in the several days prior (via urine drug screen) to week prior (via interview), and where $t$ represents the number of weeks since randomization. 
We hypothesized that
if XR-NTX reduces craving more than BUP-NX, that this could provide a
partial explanation of the lower opioid use while on XR-NTX. We use
$M_t$ to denote craving during week $t+1$ since randomization, for $t \in \{1, 2, 3\}$. There
may also be differences in depressive symptoms
\citep{rudolph2021explaining} and withdrawal symptoms \citep{tip2021}
between the two medications, so we incorporated measures of each
\citep[Hamilton Depression Scale, Subjective Opioid Withdrawal Scale;
see][]{hamilton1960hamilton, cooper2016effects}, as time-varying
confounders. We use $Z_t$ to denote these confounders measured during week $t+1$ for $t \in \{1, 2, 3\}$. 

In addition, patients could drop
out of treatment or otherwise be lost to follow-up starting in week 3 after randomization. Importantly, the outcomes $L_t$ are always observed, as it is assumed that a patient who has missing opioid use data (most likely due to a missed visit) would have been positive for opioid use \citep{hser2016long, hser2017distinctive, weiss2011adjunctive, weiss2015long}.  For $t\in\{2,3\}$, we use $A_t=1$ to denote
that a patient's $Z_t$ and $M_t$ have been measured, and we let $A_t=0$ otherwise. Thus, our intervention variable is
given by a combination of treatment and censoring,
$\bar{A}=\{A_1, A_2, A_3\}$. In summary, we can write our
observed data for this example as
$O=(L_1, A_1, Z_1, M_1, L_2, A_2, Z_2, M_2, L_3, A_3, Z_3, M_3, L_4).$

We estimated the interventional direct effect of initiating treatment
with XR-NTX vs. BUP-NX on illicit opioid use during week 4 of
treatment, not operating through craving, had those who dropped out
not dropped out. That is, for $\bar a'=(1,1,1)$ and
$\bar a^\star=(0,1,1)$, we estimated
$\E(Y(\bar a', \bar{G}(\bar a^\star)) - Y(\bar a^\star, \bar{G}(\bar
a^\star))$. We also estimated the interventional indirect effect of
initiating treatment with XR-NTX vs. BUP-NX on illicit opioid use
during week 4 of treatment operating through craving, had those who
dropped out not dropped out. That is, we estimated
$\E(Y(\bar a', \bar{G}(\bar a') - Y(\bar a', \bar{G}(\bar
a^\star))$. We used an ensemble of machine learning algorithms in
fitting the nuisance parameters. The ensemble included an
intercept-only model, lasso, multiple additive regression splines, and
extreme gradient boosting. The weights in the ensemble were chosen
using Super Learning \citep{vanderLaanPolleyHubbard07}. We used
cross-fitting with 5 folds.
 
For the interventional direct effect, we estimated that initiating
treatment with XR-NTX vs. BUP-NX, not operating through craving, would
reduce risk of using illicit opioids during week 4 of treatment by 8.8
percentage points (risk difference: -0.088, 95\% CI: -0.129,
-0.048). For the interventional indirect effect, we estimated that
initiating treatment with XR-NTX vs. BUP-NX, operating through
craving, would not meaningfully decrease risk of using illicit opioids
during week 4 of treatment (risk difference: -0.004, 95\% CI: -0.019,
0.011). Thus, we conclude that reductions in risk of illicit opioid use during treatment with XR-NTX vs. BUP-NX is not due to differences that operate
through the treatments' effects on craving.
 
\section{Discussion}

Our approach generalizes interventional causal effects to allow for
high-dimensional time-dependent variables measured post- and
pre-treatment. We present an estimation algorithm that leverages
machine learning to alleviate misspecification bias while retaining
statistical properties such as $\sqrt{n}$-consistency, non-parametric
efficiency, and asymptotic normality.

While this approach allows great flexibility in the data structure and
estimation method, some limitations remain. We assume that the
mediator is categorical, and computational tractability of our
proposed estimator requires that it takes values on a small set. This
limitation seems fundamental and hard to overcome within an
interventional effect framework, as all estimators will require
estimation of the density of the counterfactual variable
$\bar M(\bar a)$. We know of no method that can do this
non-parametrically in the case of a continuous or high-dimensional
variable $\bar M$, although recent approaches on estimation of
counterfactual densities in single time-point settings are promising
\citep{kennedy2021semiparametric}.

In addition, interventional effects have some limitations. First, they
do not decompose the average treatment effect
$E[Y(\bar a') - Y(\bar a^\star)]$. Second, the interventional indirect
effect can be non-zero, even when there is no indirect effect for any
individual in the population \citep{miles2021jsm}. Solving this
limitation would require a different framework for mediation analysis.

\bibliographystyle{plainnat}
\bibliography{refs}

\end{document}


\maketitle

\section{Auxiliary results}
The proofs of results in the paper rely heavily on existing
identification and optimality results for the longitudinal
g-computation formula. We first present those results. Consider a data
structure
$(C_1, E_1, \ldots,C_{K}, E_{K}, Y)$, Let
$\Q_{C, K+1}=Y$ and $H_{E,k} = (\bar C_k, \bar E_{k-1})$. For $k=K,\ldots,0$, let
\[\Q_{C, k}=\E[\Q_{C, k+1}\mid E_k= e_k,
  H_{E,k}, V].\] Furthermore, let
$\g_{E,k}(e_k\mid h_{E,k})= \P(E_k = e_k\mid
H_{E,k} = h_{E,k})$ and
$\G_{E,k} = \one\{E_k =
e_k\}/\g_{E,k}(e_k\mid H_{E, k})$. Define
\[\B_k = \sum_{l=k}^K \left(\prod_{u=k}^l\G_{E,
    u}\right)\{\Q_{C, l+1} - \Q_{C, l}\} + \Q_{C,k}.\]
Then we have the following three results. 
\begin{theorem}[G-computation formula. See \cite{Robins86} and \cite{Bang05}]\label{theo:gcompid}
  Let $Y(\bar e)$ denote a counterfactual outcome in a
  hypothetical world where $\bar E = \bar e$. Assume
  $Y(\bar e)\indep E_k\mid H_{E,k}$ and
  $\P(\g_{E,k}(e_k\mid H_{E,k}) > 0)=1$ for all
  $k$. Then $\E[\Q_{C, 1}]$ identifies $\E\{Y(\bar e)\}$.
\end{theorem}

\begin{theorem}[Efficient influence function. See \cite{Robins99} and
  \cite{van2012targeted}]\label{theo:eifgcomp}
  The efficient influence function for estimating
  $\E[\Q_{C, 1}]$ in the non-parametric model is given by
  $\B_1-\E[\Q_{C, 1}]$.
\end{theorem}

\begin{theorem}[Double robustness of the EIF. See \cite{Bang05} and
  \cite{luedtke2017sequential}]\label{theo:dr} For any fixed sequence
  $(\tilde
  \Q_{C, 1}, \tilde\G_{E, 1}, \ldots, \tilde
  \Q_{C, K}, \tilde\G_{E,K})$, define
  \[\tilde \C_{k,j} =
    \prod_{r=k+1}^{j-1}\tilde\G_{E,r},\]
  and
  \[\Rem_k =
    \sum_{j=k+1}^K\E\left[\tilde \C_{k,j}\{\tilde\G_{E, j} - \G_{E,
      j}\}\{\tilde \Q_{C, j} - \Q_{C, j}\}\mid
    E_k = e_k, H_{E, k}\right].\]
  Then we have
  \[\Q_{C,k} = \E[\tilde\B_{k+1}\mid E_k=e_k, H_{E,k}] +
    \Rem_k,\] where $\tilde\B_{k+1}$ is defined as above with
  $\Q_{C,k}$ and $\G_{E,k}$ replaced by
  $\tilde \Q_{C,k}$ and $\tilde\G_{E,k}$, respectively.
\end{theorem}

\section{Identification (Theorem \ref{theo:iden})}
\begin{proof}
  
  For fixed $\bar a$ and $\bar m$ we have {\small
    \begin{align}
      \E[Y&(\bar a, \bar m)\mid l_1]\notag\\
          &=\E[Y(\bar a, \bar m)\mid
            a_1, l_1]\notag\\
          &=\int \E[Y(\bar a, \bar m)\mid
            l_1, z_1,a_1]\p[z_1\mid a_1, l_1]\dd\nu(z_1)\notag\\
          &=\int \E[Y(\bar a, \bar m)\mid
            l_1, m_1, z_1, a_1]\p[z_1\mid a_1, l_1]\dd\nu(z_1)\notag\\
          &=\int\E[Y(\bar a, \bar m)\mid
            \bar l_2, m_1, z_1, a_1]\p[z_1\mid a_1, l_1]
            \p[l_2\mid  m_1, z_1, a_1]\dd\nu(l_2,z_1)\notag\\
          &=\int\E[Y(\bar a, \bar m)\mid
            \bar l_2, m_1, z_1, \bar a_2]\p[z_1\mid a_1, l_1]
            \p[l_2\mid  m_1, z_1, a_1]\dd\nu(l_2,z_1)\notag\\
          &=\int\E[Y(\bar a, \bar m)\mid
            \bar l_2, m_1, \bar z_2, \bar a_2]
            \p[l_2\mid h_{L,2}]\prod_{t=1}^2\p[z_t\mid h_{Z,t}]\dd\nu(l_2,\bar z_2)\notag\\
          &=\int\E[Y(\bar a, \bar m)\mid
            \bar l_2, \bar m_2, \bar z_2, \bar a_2]
            \p[l_2\mid h_{L,2}]\prod_{t=1}^2\p[z_t\mid h_{Z,t}]\dd\nu(l_2,\bar z_2)\notag\\
          &\vdots\notag\\
          &=\int\E[Y(\bar a, \bar m)\mid
            \bar l, \bar m, \bar z, \bar a]
            \prod_{t=2}^{\tau}\p[l_t\mid h_{L,t}]\prod_{t=1}^{\tau}\p[z_t\mid h_{Z,t}]\dd\nu(\bar l\backslash l_1,\bar z)\notag\\
          &=\int\E[Y\mid
            \bar l, \bar m, \bar z, \bar a]
            \prod_{t=2}^{\tau}\p[l_t\mid h_{L,t}]\prod_{t=1}^{\tau}\p[z_t\mid h_{Z,t}]\dd\nu(\bar l\backslash l_1,\bar z)\notag\\
          &=\int l_{\tau+1}
            \prod_{t=1}^{\tau}\p[l_{t+1}\mid h_{L,t}]\p[z_t\mid
            h_{Z,t}]\dd\nu(\bar l\backslash l_1,\bar z),\label{eq:yam}
    \end{align}}%
  where the first and fifth equalities follow from
  Assumption~\ref{ass:iden}\ref{ass:ceAY} and
  \ref{ass:pos}\ref{ass:posA}, the second, fourth, and sixth follow by
  the law of iterated expectation, and the third and seventh follow by
  Assumption~\ref{ass:iden}\ref{ass:ceMY} and
  \ref{ass:pos}\ref{ass:posM}. 

  We now identify the distribution of $M(\bar a)$. Recall we assume
  the mediator takes values on discrete set. We have: {\small
    \begin{align}
      \P[\bar M&(\bar a)= \bar m\mid l_1]=\notag\\
               &= \P[\bar M(\bar a)=\bar
                 m_\tau\mid a_1, l_1]\notag\\
               &= \int\P[\bar M(\bar a)=\bar
                 m_\tau\mid l_1, z_1, a_1]\p[z_1\mid a_1, l_1]\dd\nu(z_1)\notag\\
               &= \int\p[M_1(a_1)=m_1, \ubar M_2(\bar a)=\ubar
                 m_2\mid l_1, z_1, a_1]\p[z_1\mid a_1, l_1]\dd\nu(z_1)\notag\\
               &= \int\P[M_1=m_1, \ubar M_2(\bar a)=\ubar
                 m_2\mid l_1, z_1, a_1]\p[z_1\mid a_1,l_1]\dd\nu(z_1)\notag\\
               &= \int\P[\ubar M_2(\bar a)=\ubar
                 m_2\mid l_1,m_1, z_1, a_1]\p[m_1\mid l_1, z_1, a_1]\p[z_1\mid a_1,l_1]\dd\nu(z_1)\notag\\
               &= \int \P[\ubar M_2(\bar a)=\ubar 
                 m_2\mid \bar l_2, m_1, z_1,
                 a_1]\p[l_2\mid h_{L,2}]\p[m_1\mid h_{M,1}]\p[z_1\mid h_{Z,1}]\dd\nu(l_2,z_1)\notag\\
               &= \int \P[\ubar M_2(\bar a)=\ubar 
                 m_2\mid \bar l_2, m_1, z_1,
                 \bar a_2]\p[l_2\mid h_{L,2}]\p[m_1\mid h_{M,1}]\p[z_1\mid h_{Z,1}]\dd\nu(l_2,z_1)\notag\\
               &= \int \P[\ubar M_2(\bar a)=\ubar 
                 m_2\mid \bar l_2, m_1, \bar z_2,
                 \bar a_2]\p[l_2\mid h_{L,2}]\p[m_1\mid h_{M,1}]\prod_{t=1}^2\p[z_t\mid h_{Z,t}]\dd\nu(l_2,\bar z_2)\notag\\
               &= \int \P[\ubar M_3(\bar a)=\ubar 
                 m_3\mid \bar l_3, \bar m_2, \bar z_2,
                 \bar a_3]\prod_{t=1}^2\p[l_{t+1}\mid h_{L,t+1}]\p[z_t\mid
                 h_{Z,t}]\p[m_t\mid h_{M,t}]\dd\nu(\bar l_3\backslash l_1,\bar z_2)\notag\\
               &\vdots\notag\\
               &= \int \P[M_\tau(\bar a)=
                 m_\tau\mid a_\tau, h_{A,\tau}]\prod_{t=1}^{\tau-1}\p[l_{t+1}\mid h_{L,t+1}]\p[z_t\mid
                 h_{Z,t}]\p[m_t\mid h_{M,t}]\dd\nu(\bar
                 l_\tau\backslash l_1,\bar
                 z_{\tau-1})\notag\\
               &= \int \P[M_\tau=
                 m_\tau\mid a_\tau, h_{A,\tau}]\prod_{t=1}^{\tau-1}\p[l_{t+1}\mid h_{L,t+1}]\p[z_t\mid
                 h_{Z,t}]\p[m_t\mid h_{M,t}]
                 \dd\nu(\bar
                 l_\tau\backslash l_1,\bar
                 z_{\tau-1})\notag\\
               &= \int \prod_{t=1}^\tau\p[l_{t+1},m_t,z_t\mid
                 h_{Z,t}]\dd\nu(\bar l\backslash l_1,\bar z)
                 \label{eq:ma}
    \end{align}
  }%
  where the first, seventh, and ninth equalities follow from
  Assumption~\ref{ass:iden}\ref{ass:ceAM} and
  \ref{ass:pos}\ref{ass:posA}, the second, sixth, and eight follow by
  the law of iterated expectation, the third and fourth by definition
  of $\bar M(\bar a)$ and the NPSEM, and the fifth by Bayes theorem.

  Since $\bar G(\bar a^\star)$ is distributed as
  $\bar M(\bar a^\star)$ and is independent of all data, we have
  \begin{align*}
    \E[Y(\bar a', \bar G(\bar a^\star))]&=\int \E[Y(\bar a',
                                          \bar m)\mid \bar
                                          G(\bar a^\star) = \bar m]
                                          \P[\bar M(\bar a)= \bar
                                                     m]\dd\nu(\bar
                                                     m)\\
                                        &=\int \E[Y(\bar a',
                                          \bar m)]
                                          \P[\bar M(\bar a^\star)= \bar
                                          m]\dd\nu(\bar
                                          m).
  \end{align*}
  Marginalizing expressions (\ref{eq:yam}) and (\ref{eq:ma}) over
  $l_1$ and plugging in the resulting expressions in the above formula
  concludes the proof of the theorem.
\end{proof}

\section{Sequential regression representation (Proposition~\ref{prop:rep})}

\begin{proof}
  In this proof, we will simplify the formula of
  Theorem~\ref{theo:iden} using the following definitions for
  $u=\tau,\ldots, 1$:
  \begin{align*}
    \varphi_{L,u}(\ubar m_u, h_{M,u}') &= \int l_{\tau+1}
    \prod_{t=u}^{\tau}\p[l_{t+1}\mid h_{L,t}']\p[z_t\mid
    h_{Z,t}']\dd\nu(\bar l_u,\bar z_{u+1}')\\
    \varphi_{Z,u}(\ubar m_u, h_{A,u}') &= \int l_{\tau+1}
    \prod_{t=u}^{\tau}\p[l_{t+1}\mid h_{L,t}']\p[z_t\mid
    h_{Z,t}']\dd\nu(\bar l_u,\bar z_u),
  \end{align*}
  where we omit the dependence of the above functions on $\ubar a_u'$
  to simplify notation.  We will first prove that
  $\varphi_{L,0}(\bar m)=\Q_{L,0}(\bar m)$, using induction.  First,
  note that
  \begin{align*}
    \varphi_{L,u}(\ubar m_u, h_{M,u}') &= \int \varphi_{Z,u+1}(\ubar m_{u+1},  h_{A,u+1})\dd\P[l_{u+1}\mid h_{L,u+1}']\\
    \varphi_{Z,u}(\ubar m_u, h_{A,u}') &= \int \varphi_{L,u}(\ubar m_u,h_{M,u}')\dd\P[z_u\mid
                                       h_{Z,t}'].
  \end{align*}
For $u=\tau$, we have
\begin{align}
  \Q_{L,u} &= \E[Y\mid  M_u = m_u, H_{M,u}]\notag\\
           &=\int Y \dd\P(Y\mid M_u=m_u,H_{M,u})\notag\\
           &= \varphi_{L,u}(\ubar m_u, H_{M,u})\notag\\
  \Q_{Z,u} &= \E[\Q_{L,u}\mid  A_u =
             a_u' ,H_{A,u}]\notag\\
           &=\int \varphi_{L,u}(\ubar m_u, H_{M,u})\dd\P(Z_u\mid A_u=a_u', 
             H_{A,u})\notag\\
           &=\varphi_{Z,u}(\ubar m_u,H_{A,u}),\label{eq:qz}
\end{align}
We've shown that $\Q_{Z,u}=\varphi_{Z,u}(\ubar
m_u,H_{A,u})$ and $\Q_{L,u}=\varphi_{L,u}(\ubar
m_u,H_{M,u})$ for
$u=\tau$. We will now assume that these equalities holds for any
$u+1$, and will prove that they hold for $u$. We have
\begin{align*}
  \Q_{L,u} &= \E[\Q_{Z,u+1}\mid  M_u = m_u, H_{M,u}]\\
           &=\int \varphi_{Z,u+1}(\ubar m_{u+1},H_{A,u+1}) \dd\P(L_{u+1}\mid M_u = m_u, H_{M,u})\\
           &= \varphi_{L,u}(\ubar m_u, H_{M,u}),
\end{align*}
where the second equality holds by assumption, and the last by definition of
$\varphi_{L,u}$. Notice that (\ref{eq:qz}) holds for any $u$ by
assumption. This completes the inductive argument to show that
$\varphi_{L,0}(\bar m)=\varphi(\bar m)=\Q_{L,0}(\bar m)$.

We will now show that $\lambda(\bar m) = \Q_{M,0}(\bar m)$. Let
$\Q^\diamond_{M,\tau+1}(\bar m) = \one\{\bar M = \bar m\}$. For
$t=\tau,\ldots, 1$, define
\[\Q^\diamond_{M,t}(\bar m)= \E[\Q^\diamond_{M,t+1}(\bar m)\mid
  A_t=a_t^\star, H_{A,t}].\] Then, the g-formula of \cite{Robins86}
(Theorem~\ref{theo:gcompid} above) shows that
$\Q^\diamond_{M,0}(\bar m) =\lambda(\bar m)$. We now prove that
$\Q^\diamond_{M,0}(\bar m) =\Q_{M,0}(\bar m)$. At $t=\tau$, we have
\begin{align*}
  \Q^\diamond_{M,t}(\bar m)&=\E[\one\{\bar M_t=\bar m_t\}\mid A_t=a_t^\star, H_{A,t}]\\&=\prod_{s=1}^{t-1}\one\{M_s=
  m_s\}\E[\one\{M_t=m_t\}\mid A_t=a_t^\star, H_{A,t}]\\
  &=\prod_{s=1}^{t-1}\one\{M_s=
  m_s\}\Q_{M,t}(\ubar m_t),
  \end{align*}
This yields
\begin{align*}
  \Q^\diamond_{M,t-1}(\bar m)
  &=\E[\Q^\diamond_{M,t}(\bar m)\mid A_{t-1}=a_{t-1}^\star, H_{A,t-1}]\\
  &=\E\left[\prod_{s=1}^{t-1}\one\{M_s=
    m_s\}\Q_{M,t}(\ubar m_t)\mid A_{t-1}=a_{t-1}^\star,
    H_{A,t-1}\right]\\
  &=\prod_{s=1}^{t-2}\one\{M_s=
    m_s\}\E\left[\one\{M_{t-1}=m_{t-1}\}\Q_{M,t}(\ubar m_t)\mid A_{t-1}=a_{t-1}^\star,
    H_{A,t-1}\right]\\
  &=\prod_{s=1}^{t-2}\one\{M_s=
    m_s\}\Q_{M,t-1}(\ubar m_{t-1}).
\end{align*}
Sequential application of this expression yields
\[\Q^\diamond_{M,t}(\bar m) = \prod_{s=1}^{t-1}\one\{M_s=
  m_s\}\Q_{M,t}(\ubar m_t),\] for $t=\tau,\ldots,1$. An
application to $t\leq 1$ yields the desired result.
\end{proof}

\section{First order approximation of parameter functionals (Lemma~\ref{lemma:vm})}
For each $t=0,\ldots,\tau$ and for any fixed $\tilde\eta$, define
the second order terms {\footnotesize
  \begin{align*}
    \Rem_{Z, t}(\eta,\tilde\eta)
    & = \sum_{s=t}^\tau\E\left[\tilde \C_{t+1,s}'\tilde \H_{t,s-1}\{\tilde\G_{M,s} - \G_{M,s}\}\{\tilde
      \Q_{L,s}-\Q_{L,s}\}\mid  A_t=a_t', H_{A,t}\right]\\
    &+\sum_{s=t+1}^\tau\E\left[\tilde \C_{t+1,s-1}'\tilde \H_{t,s-1}\{\tilde\G_{A,s}' - \G_{A,s}'\}\{\tilde
      \Q_{Z,s}-\Q_{Z,s}\}\mid \ubar A_t=a_t', H_{A,t}\right]\\
    \Rem_{L, t}(\eta,\tilde\eta)
    &=\sum_{s=t+1}^\tau\E\left[\tilde \C_{t+1,s}'\tilde \H_{t+1,s-1}\{\tilde\G_{M,s} - \G_{M,s}\}\{\tilde
      \Q_{L,s}-\Q_{L,s}\}\mid M_t=m_t, H_{M,t}\right]\\
    &+\sum_{s=t+1}^\tau\E\left[\tilde \C_{t+1,s-1}'\tilde \H_{t+1,s-1}\{\tilde\G_{A,s}'- \G_{A,s}'\}\{\tilde
      \Q_{Z,s}-\Q_{Z,s}\}\mid M_t=m_t, H_{M,t}\right]\\
    \Rem_{M,t}(\eta, \tilde\eta)
    &=\sum_{s=t+1}^\tau\E\left[\tilde \C_{t+1,s-1}^\star\{\tilde\G_{A,s} - \G_{A,s}\}\{\tilde
      \Q_{M,s}-\Q_{M,s}\}\mid A_t  = a_t^\star, H_{A,t}\right].
  \end{align*}
}%

\begin{proof}
In this proof we use the notation $\Q_{Z,t}(\bar m)$, $\Q_{L,t}(\bar m)$, $\D_{Z,t}(\bar
m, \tilde \eta)$ and $\D_{L,t}(\bar
m, \tilde \eta)$ to make explicit the dependence
of these random variables on $\bar m$. We will first prove that
the expansions for $\Q_{Z,t}(\bar m)$ and $\Q_{L,t}(\bar m)$ hold.

For $t=1,\ldots,\tau$, let $C_{2t-1} = L_{i,t}$, $E_{2t-1} = A_t$,
$C_{2t} = Z_{i,t}$, $E_{2t} = M_t$.
\begin{table}[!htb]
  \centering   
  \begin{tabular}[H]{c|c|c|c|c|c|c|c|c|c|c}
    $L_1$ & $A_1$ & $Z_1$ & $M_1$ & $\ldots$ & $L_t$ & $A_t$ & $Z_t$ &  $M_t$ &$\ldots$ & $Y$\\
    $C_1$ & $E_1$ & $C_2$ & $E_2$ & $\ldots$ & $C_{2t-1}$ & $E_{2t-1}$ & $C_{2t}$ & $E_{2t}$ &$\ldots$ & $Y$\\
    $\Q_{Z,1}$ & $\G_{A,1}$ & $\Q_{L,1}$ & $\G_{M,1}$ & $\ldots$ & $\Q_{Z,t}$ & $\G_{A,t}$ & $\Q_{L,t}$ & $\G_{M,t}$ & $\ldots$ & $Y$\\
    $\Q_{C,1}$ & $\G_{E,1}$ & $\Q_{C,2}$ & $\G_{E,2}$ & $\ldots$ & $\Q_{C,2t-1}$ & $\G_{E,2t-1}$ & $\Q_{C,2t}$ & $\G_{E,2t}$ & $\ldots$ &$Y$\\
  \end{tabular}
  \caption{Mapping between different data structures}
  \label{tab:map}
\end{table}Then the data structure of Theorem~\ref{theo:dr} is
equivalent to
$X_i=(L_{i,1}, A_{i,1}, Z_{i,1}, M_{i,1}, \ldots, \allowbreak
L_{i,\tau}, A_{i,\tau}, Z_{i,\tau}, M_{i,\tau}, Y_i)$ with
$K=2\tau$. Then, Lemma~\ref{lemma:vm} follows from
Theorem~\ref{theo:dr} applied to this data structure with an
intervention equal to $e_{2t-1}= a_t'$ and $e_{2t}= m_t$ for
$t=1,\ldots,\tau.$ See Table~\ref{tab:map} for a mapping of the
various quantities between these two data structures. For even $k$,
Theorem~\ref{theo:dr} applied to this data structure proves that
\[  \Q_{Z,t}(\bar m) 
  =\E[\D_{L,t}(\bar m, \tilde \eta)\mid
     A_t=a_t', H_{A,t}] +
     \Rem_{Z, t}(\bar m, \tilde \eta),\]
   and for odd $k$ it shows that
   \[ \Q_{L,t}(\ubar m_t)
     =\E[\D_{Z,t+1}(\bar m, \tilde \eta) \mid M_t = m_t, H_{M,t}] +
     \Rem_{L, t}(\bar m, \tilde \eta).\] The proof for $\bar Q_{M,t}$
   proceeds as follows.  For $t =\tau+1$, define
   $\Q^\diamond_{M,t} = \one\{\bar M = \bar m\}$, as before. For
   $t=\tau,\ldots, 1$, recursively define
  \[\Q^\diamond_{M,t} = \E[\Q^\diamond_{M,t+1}\mid A_t = a_t^\star,
    H_{A,t}].\] Theorem~\ref{theo:gcompid} with $\tau=K$, $E_k=A_t$,
  $C_k=L_t$, and $Y=\one\{\bar M = \bar m\}$ shows that
  \begin{equation}
    \Q^\diamond_{M,t}
    =\E[\D^\diamond_{M,t+1}(\tilde \eta)
    \mid A_t=a_t^\star, H_{A,t}] +
     \Rem^\diamond_{M,t}(\tilde \eta),\label{eq:checkM}
  \end{equation}
  where
  \[ \Rem^\diamond_{M,t}(\tilde \eta) = \sum_{s=t+1}^\tau\E\left[\tilde \C_{t,s}\{\tilde\G_{A,s} - \G_{A,s}\}\{\tilde
      \Q^\diamond_{M,s}-\Q^\diamond_{M,s}\}\mid A_t  = a_t^\star,
      H_{A,t}\right],\]
  and
  \[\D^\diamond_{M,t}(\tilde\eta)= \sum_{s=t}^\tau\left(\prod_{u=t}^s\tilde \G_{A,u}\right)
    \left\{\tilde\Q^\diamond_{M,s+1} - \tilde\Q^\diamond_{M,s}\right\}
    +  \tilde\Q^\diamond_{M,t}.\]
  Replacing $\tilde\Q^\diamond_{M,s} = \prod_{k=1}^{s-1}\one\{M_k=
  m_k\}\tilde\Q_{M,s}(\ubar m_s)$ in the above expression and some
  algebra yields
  \begin{align*}
    &\D^\diamond_{M,t}(\ubar X_t, \bar m,\tilde\eta)\\
    &=\sum_{s=t}^\tau\left(\prod_{u=t}^s\tilde \G_{A,u}\right)
      \left\{\prod_{k=1}^{s}\one\{M_k=
      m_k\}\tilde\Q_{M,s+1} - \prod_{k=1}^{s-1}\one\{M_k=
      m_k\}\tilde\Q_{M,s}\right\}
      +  \prod_{k=1}^{t-1}\one\{M_k=
      m_k\}\tilde\Q_{M,t}\\
    &=\sum_{s=t}^\tau\left(\prod_{u=t}^s\tilde \G_{A,u}\right)\prod_{k=1}^{s-1}\one\{M_k=
      m_k\}
      \left\{\one\{M_s=
      m_s\}\tilde\Q_{M,s+1} - \tilde\Q_{M,s}\right\}
      +  \prod_{k=1}^{t-1}\one\{M_k=
      m_k\}\tilde\Q_{M,t}\\
    &=\prod_{k=1}^{t-1}\one\{M_k=
      m_k\}\left[\sum_{s=t}^\tau\left(\prod_{u=t}^s\tilde \G_{A,u}\right)\prod_{k=t}^{s-1}\one\{M_k=
      m_k\}
      \left\{\one\{M_s=
      m_s\}\tilde\Q_{M,s+1} - \tilde\Q_{M,s}\right\}
      +  \tilde\Q_{M,t}\right]\\    
  &=\prod_{k=1}^{t-1}\one\{M_k= m_k\}\D_{M,t}(\ubar X_t, \ubar
    m_t;\tilde\eta),
  \end{align*}
which in turn shows that
\begin{align*}
  \prod_{k=1}^{t-1}&\one\{M_k=
  m_k\}\Q_{M,t}(\ubar m_t)= \\
  &=\Q^\diamond_{M,t}(\bar m)\\
  &=\E[\D^\diamond_{M,t+1}(\tilde \eta)
  \mid A_t=a_t^\star, H_{A,t}] +
    \Rem_{M,t}(\tilde \eta)\\
  &=\E\left[\prod_{k=1}^t\one\{M_k= m_k\}\D_{M,t+1}(\ubar X_{t+1}, \ubar
    m_{t+1};\tilde\eta)
    \mid A_t=a_t^\star, H_{A,t}\right]   +
    \Rem^\diamond_{M,t}(\tilde \eta)\\
  &=\prod_{k=1}^{t-1}\one\{M_k= m_k\}\E\left[\one\{M_t=m_t\}\D_{M,t+1}(\ubar X_{t+1}, \ubar
    m_{t+1};\tilde\eta)
    \mid A_t=a_t^\star, H_{A,t}\right]   +
    \Rem^\diamond_{M,t}(\tilde \eta).
\end{align*}
Finally, note that if $\bar M_{t-1}=\bar m_{t-1}$, then $\Rem^\diamond_{M,t}(\tilde \eta)=\Rem_{M,t}(\tilde \eta)$.
\end{proof}

\section{Efficient influence function (Theorem~\ref{theo:eif})}
\begin{proof}
  In this proof we will show that the EIF for $\varphi(\bar m)$ is
  equal to $\D_{Z,1}(X,\bar m;\eta) - \varphi(\bar m)$, and that the
  EIF for $\lambda(\bar m)$ is
  equal to $\D_{M,1}(X,\bar m;\eta) - \lambda(\bar m)$. The theorem
  then follows from an application of the Delta method to the result
  in Proposition~\ref{prop:rep}.

  We will prove the result for $\varphi(\bar m)$, analogous arguments
  yield the result for $\lambda(\bar m)$. For a parameter
  $\varphi:\P\to \R$, and for a submodel $\P_\epsilon$ such that
  $\P_{\epsilon=0} = \P$, the EIF is defined as the unique function
  $\D$ such that
  \[\frac{\dd\varphi(\P_\epsilon)}{\dd\epsilon}\bigg|_{\epsilon
      =0}=\E[\D(X)s(X)],\]
  with $s(X)$ equal to the score function of the submodel
  $\P_\epsilon$. For $t=0$, Lemma~\ref{lemma:vm} teaches us that for
  any two distributions $\F_1$ and $\F_2$ in the model (corresponding
  to nuisance parameters $\tilde\eta$ and $\eta$), we have:
  \[\varphi(\F_1) - \varphi(\F_2) = \int \D_{Z,1}(x;\F_2)\dd\F_1(x) +
    \Rem_{Z,0}(\F_1,\F_2),\]
  where $\Rem_{M,0}(\F_1,\F_2)$ is a second order term of the type
  \[\Rem_{M,0}(\F_1,\F_2) = \sum_{s=1}^\tau \int
    c(x,\F_2)\{[f_s(x,\F_1)-f_s(x,\F_2)][g_s(x,\F_1)-g_s(x,\F_2)]\}\dd\F_1(x)\]
  for transformations $f_s$ and $g_s$.  Evaluating the above expression
  at $\F_1=\P_\epsilon$ and $\F_2=\P$, differentiating with respect to
  $\epsilon$, and evaluating at $\epsilon=0$ yields
  \begin{align*}
    \frac{\dd }{\dd \epsilon}\varphi(\P_\epsilon)\bigg|_{\epsilon = 0}
    &= \int \D_{Z,1}(\cdot;\P)
      \left(\frac{\dd  \P_\epsilon}{\dd \epsilon}\right)\bigg|_{\epsilon
      = 0} + \frac{\dd }{\dd
      \epsilon}\Rem_{Z,0}(\P_\epsilon,\P)\bigg|_{\epsilon = 0} \\
    &= \int\D_{Z,1}(\cdot;\P)
      \left(\frac{\dd \log \P_\epsilon}{\dd
      \epsilon}\right)\bigg|_{\epsilon = 0}\dd \P + \frac{\dd }{\dd
      \epsilon}\Rem_{Z,0}(\P_\epsilon,\P) \bigg|_{\epsilon = 0},
  \end{align*}
  and the result follows after noticing that
  \[\frac{\dd }{\dd \epsilon}\Rem_{Z,0}(\P_\epsilon,\P)\bigg|_{\epsilon = 0}=0.\]
\end{proof}

\section{Theorem~\ref{theo:weak}}
\begin{proof}
  First, note that an application of Theorem 4 of
  \cite{diaz2021nonparametricmtp} to the augmented data structure in
  Table~\ref{tab:map} yields the weak convergence of
  $\hat\varphi(\hat m)$ and $\hat\lambda(\hat m)$. The theorem then
  follows by an application of the Delta method.
\end{proof}
\section{Counterfactual definition in terms of the assumed NPSEM}
Let $\bar M(\bar a)$ denote the counterfactual mediator vector
observed in a hypothetical world where $\bar A = \bar a$. Each entry
of this vector is defined as
$M_t(\bar a_t) = f_{M,t}(H_{M,t}(\bar a_t), U_{M,t})$, where the
counterfactual history is defined recursively as
\begin{align*}
  H_{Z,t}(\bar a_t) &= (\bar L_t(\bar a_{t-1}), \bar M_{t-1}(\bar a_{t-1}), \bar
                      Z_{t-1}(\bar a_{t-1}), \bar a_t),\\
  H_{M,t}(\bar a_t) &= (\bar L_t(\bar a_{t-1}), \bar M_{t-1}(\bar a_{t-1}), \bar
                      Z_t(\bar a_t), \bar a_t),\\
  H_{L,t}(\bar a_{t-1}) &= (\bar L_{t-1}(\bar a_{t-2}), \bar M_{t-1}(\bar a_{t-1}), \bar
                          Z_{t-1}(\bar a_{t-1}), \bar a_{t-1}),
\end{align*}
with
\begin{align*}
  Z_t(\bar a_t) &= f_{Z,t}(H_{Z,t}(\bar a_t), U_{Z,t}),\\
  M_t(\bar a_t) &= f_{M,t}(H_{M,t}(\bar a_t), U_{M,t}),\\
  L_t(\bar a_{t-1}) &= f_{L,t}(H_{L,t}(\bar a_{t-1}), U_{L,t}).
\end{align*}
For a value $\bar m$ in the range of $\bar M$ we also define a
counterfactual outcome $Y(\bar a, \bar m)= L_{\tau+1}(\bar a, \bar m) =
f_{L,\tau+1}(H_{L,\tau+1}(\bar a, \bar m), U_{L,\tau+1})$, where
\begin{align*}
  H_{Z,t}(\bar a_t, \bar m_{t-1}) &= (\bar L_t(\bar a_{t-1}, \bar m_{t-1}), \bar m_{t-1}, \bar
                                    Z_{t-1}(\bar a_{t-1}, \bar m_{t-2}), \bar a_t),\\
  H_{L,t}(\bar a_{t-1}, \bar m_{t-1}) &= (\bar L_{t-1}(\bar a_{t-2}, \bar m_{t-2}), \bar m_{t-1}, \bar
                                        Z_{t-1}(\bar a_{t-1}, \bar m_{t-2}), \bar a_{t-1}),
\end{align*}
with
\begin{align*}
  Z_t(\bar a_t, \bar m_t) &= f_{Z,t}(H_{Z,t}(\bar a_t, \bar m_{t-1}), U_{Z,t}),\\
  L_t(\bar a_{t-1}, \bar m_{t-1}) &= f_{L,t}(H_{L,t}(\bar a_{t-1}, \bar m_{t-1}), U_{L,t}).
\end{align*}



\section{Baseline covariates used for illustrative application}
Baseline covariates included study site, gender, age, homeless status, race/ethnicity, level of education, employment, history of IV drug use, duration of opioid use,
alcohol use disorder, cocaine use disorder, current use of: heroin, methadone, amphetamines, sedatives, cannabis, cocaine; current methadone prescription, current smoker, whether or not lives with someone with problematic alcohol use, whether or not lives with someone who uses drugs, whether or not this was first treatment for OUD, any friends/family with: alcohol problems, heroin or other illicit opioid use, or other illicit drug use; history of psychiatric disorder, prior opioid withdrawal discomfort level, baseline pain score, baseline depressive score, severity of opioid use, and timing of randomization. 

\bibliographystyle{plainnat} \bibliography{refs}

%% file: _preamble.tex
\usepackage[inline,shortlabels]{enumitem}
\usepackage{float}
\usepackage{mathtools}
\usepackage{subdepth}
\usepackage{graphicx}
\usepackage[round]{natbib}
\usepackage[margin=1in]{geometry}
\usepackage{tikz}
\usetikzlibrary{arrows.meta}
\usetikzlibrary{decorations.markings}
\usepackage[linesnumbered]{algorithm2e}
\RestyleAlgo{ruled}

\tikzset{+ /.tip = {Bar[sep=-3pt 2,width=3pt 4]_[sep=0]}}
\usepackage[english]{babel}
\usepackage{longtable}
\usepackage{color}
\usepackage{mathtools}
\usepackage{amssymb,amsmath,amsthm}
\usepackage{multirow}
\usepackage[titletoc,title]{appendix}
\usepackage{authblk}
\usepackage{bbm}
\usepackage{setspace}
\usepackage{dsfont}
\usepackage[OT1]{fontenc}
\usepackage{subcaption}
\usepackage{refcount}
\usepackage{booktabs}
\graphicspath{ {../simulations/01_init_manuscript/graphs/} }

\usepackage{xr-hyper}
\usepackage[colorlinks,citecolor=blue,urlcolor=blue]{hyperref}

\usepackage[inline]{trackchanges}
\addeditor{Ivan}
\addeditor{Nima}

\newtheorem{prop}{Proposition}
\newtheorem{theorem}{Theorem}
\AtEndDocument{\refstepcounter{theorem}\label{finalthm}}
\AtEndDocument{\refstepcounter{equation}\label{finaleq}}

{
  \theoremstyle{definition}
  
}
{
  \theoremstyle{definition}
  \newtheorem{assumptioniden}{}
}
{
  \theoremstyle{definition}
  
}

\newtheorem{lemma}{Lemma}
\AtEndDocument{\refstepcounter{lemma}\label{finallemma}}

 \DeclareMathOperator{\var}{\mathsf{Var}}

\renewcommand{\P}{\mathsf{P}}
\newcommand{\p}{\mathsf{p}}
\newcommand{\Q}{\mathsf{Q}}

\newcommand{\Rem}{\mathsf{R}}
\newcommand{\g}{\mathsf{g}}

\newcommand{\G}{\mathsf{G}}
\let\sec\S 

\renewcommand{\S}{\mathsf{S}}
\newcommand{\C}{\mathsf{K}}
\renewcommand{\H}{\mathsf{H}}

\newcommand{\indep}{\mbox{$\perp\!\!\!\perp$}} 
 \newcommand{\dd}{\,\mathrm{d}}
\newcommand{\Pn}{\mathsf{P}_n}

\newcommand{\D}{\mathsf{D}}
\newcommand{\IF}{\mathsf{IF}}

\newcommand{\one}{\mathds{1}}
 
\newcommand{\E}{\mathsf{E}}
\renewcommand{\P}{\mathsf{P}}

\usepackage{accents}

\newcommand{\ubar}[1]{\underaccent{\bar}{#1}}

\DeclarePairedDelimiterX{\norm}[1]{\lVert}{\rVert}{#1}

\pgfdeclarelayer{background}
\pgfsetlayers{background,main}
\usetikzlibrary{arrows,positioning}
\tikzset{
>=stealth',
punkt/.style={
rectangle,
rounded corners,
draw=black, very thick,
text width=6.5em,
minimum height=2em,
text centered},
pil/.style={
->,
thick,
shorten <=2pt,
shorten >=2pt,}
}
\newcommand{\Vertex}[3]
{\node[minimum width=0.6cm,inner sep=0.05cm] (#2) at (#1) {#3};
}
\newcommand{\Vertexr}[3]
{\node[rectangle, draw, minimum width=0.6cm,inner sep=0.05cm] (#2) at (#1) {#2};
}
\newcommand{\ArrowR}[3]%
{ \begin{pgfonlayer}{background}
\draw[->,#3] (#1) to[bend right=30] (#2);
\end{pgfonlayer}
}
\newcommand{\ArrowLW}[3]%
{ \begin{pgfonlayer}{background}
\draw[->,#3] (#1) to[bend left=30] (#2);
\end{pgfonlayer}
}

\newcommand{\ArrowL}[3]%
{ \begin{pgfonlayer}{background}
    \draw[->,#3] (#1) to[bend left=45] (#2);
  \end{pgfonlayer}
}

\newcommand{\EdgeL}[3]%
{ \begin{pgfonlayer}{background}
\draw[dashed,#3] (#1) to[bend right=-45] (#2);
\end{pgfonlayer}
}

\newcommand{\Arrow}[3]%
{ \begin{pgfonlayer}{background}
\draw[->,#3] (#1) -- +(#2);
\end{pgfonlayer}
}

\allowdisplaybreaks
\pgfarrowsdeclare{arcs}{arcs}{...}
{
  \pgfsetdash{}{0pt} 
  \pgfsetroundjoin   
  \pgfsetroundcap    
  \pgfpathmoveto{\pgfpoint{-5pt}{5pt}}
  \pgfpatharc{180}{270}{5pt}
  \pgfpatharc{90}{180}{5pt}
  \pgfusepathqstroke
}
\newcommand{\ArrowB}[3]%
{ \begin{pgfonlayer}{background}
    \draw[|-arcs,line width=0.4mm,shorten <= 0.3cm,shorten >= 0.3cm,#3] (#1) -- +(#2);
  \end{pgfonlayer}
}

\newcommand{\titlepaper}{Efficient and flexible causal mediation with
  time-varying mediators, treatments, and confounders}

\date{\today}

\author[1]{Iv\'an D\'iaz \thanks{corresponding author:
    ild2005@med.cornell.edu}}
\author[2]{Nicholas Williams}
\author[2]{Kara E. Rudolph} 
\affil[1]{\small Division of Biostatistics, Weill Cornell Medicine.}
\affil[2]{\small Department of
   Epidemiology, Mailman School of Public Health, Columbia University.}

%% file: paper_arxiv.bbl
\begin{thebibliography}{63}
\providecommand{\natexlab}[1]{#1}
\providecommand{\url}[1]{\texttt{#1}}
\expandafter\ifx\csname urlstyle\endcsname\relax
  \providecommand{\doi}[1]{doi: #1}\else
  \providecommand{\doi}{doi: \begingroup \urlstyle{rm}\Url}\fi

\bibitem[Andrews and Didelez(2020)]{andrews2020insights}
Ryan~M Andrews and Vanessa Didelez.
\newblock Insights into the cross-world independence assumption of causal
  mediation analysis.
\newblock \emph{Epidemiology}, 32\penalty0 (2):\penalty0 209--219, 2020.

\bibitem[Bang and Robins(2005)]{Bang05}
Heejung Bang and James~M Robins.
\newblock Doubly robust estimation in missing data and causal inference models.
\newblock \emph{Biometrics}, 61\penalty0 (4):\penalty0 962--973, 2005.

\bibitem[Benkeser and van~der Laan(2016)]{benkeser2016highly}
David Benkeser and Mark van~der Laan.
\newblock The highly adaptive lasso estimator.
\newblock In \emph{2016 IEEE International Conference on Data Science and
  Advanced Analytics (DSAA)}, pages 689--696. IEEE, 2016.

\bibitem[Bickel(1982)]{bickel1982adaptive}
Peter~J Bickel.
\newblock On adaptive estimation.
\newblock \emph{The Annals of Statistics}, pages 647--671, 1982.

\bibitem[Bickel et~al.(1997)Bickel, Klaassen, Ritov, and Wellner]{Bickel97}
Peter~J Bickel, Chris~AJ Klaassen, YA'Acov Ritov, and Jon~A Wellner.
\newblock \emph{Efficient and Adaptive Estimation for Semiparametric Models}.
\newblock Springer-Verlag, 1997.

\bibitem[Bickel et~al.(2009)Bickel, Ritov, Tsybakov,
  et~al.]{bickel2009simultaneous}
Peter~J Bickel, Ya’acov Ritov, Alexandre~B Tsybakov, et~al.
\newblock Simultaneous analysis of lasso and dantzig selector.
\newblock \emph{The Annals of Statistics}, 37\penalty0 (4):\penalty0
  1705--1732, 2009.

\bibitem[Bind et~al.(2016)Bind, Vanderweele, Coull, and
  Schwartz]{bind2016causal}
M-AC Bind, TJ~Vanderweele, BA~Coull, and JD~Schwartz.
\newblock Causal mediation analysis for longitudinal data with exogenous
  exposure.
\newblock \emph{Biostatistics}, 17\penalty0 (1):\penalty0 122--134, 2016.

\bibitem[Chen and White(1999)]{chen1999improved}
Xiaohong Chen and Halbert White.
\newblock Improved rates and asymptotic normality for nonparametric neural
  network estimators.
\newblock \emph{IEEE Transactions on Information Theory}, 45\penalty0
  (2):\penalty0 682--691, 1999.

\bibitem[Chernozhukov et~al.(2018)Chernozhukov, Chetverikov, Demirer, Duflo,
  Hansen, Newey, and Robins]{chernozhukov2018double}
Victor Chernozhukov, Denis Chetverikov, Mert Demirer, Esther Duflo, Christian
  Hansen, Whitney Newey, and James Robins.
\newblock Double/debiased machine learning for treatment and structural
  parameters.
\newblock \emph{The Econometrics Journal}, 21\penalty0 (1):\penalty0 C1--C68,
  2018.

\bibitem[Cooper et~al.(2016)Cooper, Johnson, Pavlicova, Glass, Vosburg,
  Sullivan, Manubay, Martinez, Jones, Saccone, et~al.]{cooper2016effects}
Ziva~D Cooper, Kirk~W Johnson, Martina Pavlicova, Andrew Glass, Suzanne~K
  Vosburg, Maria~A Sullivan, Jeanne~M Manubay, Diana~M Martinez, Jermaine~D
  Jones, Phillip~A Saccone, et~al.
\newblock The effects of ibudilast, a glial activation inhibitor, on opioid
  withdrawal symptoms in opioid-dependent volunteers.
\newblock \emph{Addiction biology}, 21\penalty0 (4):\penalty0 895--903, 2016.

\bibitem[D{\'\i}az and Hejazi(2020)]{diaz2020causal}
Iv{\'a}n D{\'\i}az and Nima~S Hejazi.
\newblock Causal mediation analysis for stochastic interventions.
\newblock \emph{Journal of the Royal Statistical Society: Series B (Statistical
  Methodology)}, 82\penalty0 (3):\penalty0 661--683, 2020.

\bibitem[D{\'\i}az et~al.(2021{\natexlab{a}})D{\'\i}az, Hejazi, Rudolph, and
  van Der~Laan]{diaz2021nonparametric}
Iv{\'a}n D{\'\i}az, Nima~S Hejazi, Kara~E Rudolph, and Mark~J van Der~Laan.
\newblock Nonparametric efficient causal mediation with intermediate
  confounders.
\newblock \emph{Biometrika}, 108\penalty0 (3):\penalty0 627--641,
  2021{\natexlab{a}}.

\bibitem[D{\'\i}az et~al.(2021{\natexlab{b}})D{\'\i}az, Williams, Hoffman, and
  Schenck]{diaz2021nonparametricmtp}
Iv{\'a}n D{\'\i}az, Nicholas Williams, Katherine~L Hoffman, and Edward~J
  Schenck.
\newblock Nonparametric causal effects based on longitudinal modified treatment
  policies.
\newblock \emph{Journal of the American Statistical Association}, pages 1--16,
  2021{\natexlab{b}}.

\bibitem[Didelez(2019)]{didelez2019defining}
Vanessa Didelez.
\newblock Defining causal mediation with a longitudinal mediator and a survival
  outcome.
\newblock \emph{Lifetime data analysis}, 25\penalty0 (4):\penalty0 593--610,
  2019.

\bibitem[Gilbert et~al.(2021)Gilbert, Montefiori, McDermott, Fong, Benkeser,
  Deng, Zhou, Houchens, Martins, Jayashankar, et~al.]{gilbert2021immune}
Peter~B Gilbert, David~C Montefiori, Adrian McDermott, Youyi Fong, David~C
  Benkeser, Weiping Deng, Honghong Zhou, Christopher~R Houchens, Karen Martins,
  Lakshmi Jayashankar, et~al.
\newblock Immune correlates analysis of the mrna-1273 covid-19 vaccine efficacy
  trial.
\newblock \emph{MedRxiv}, 2021.

\bibitem[Greiner et~al.(2021)Greiner, Shulman, Choo, Scodes, Pavlicova,
  Campbell, Novo, Fishman, Lee, Rotrosen, et~al.]{greiner2021naturalistic}
Miranda~G Greiner, Matisyahu Shulman, Tse-Hwei Choo, Jennifer Scodes, Martina
  Pavlicova, Aimee~NC Campbell, Patricia Novo, Marc Fishman, Joshua~D Lee, John
  Rotrosen, et~al.
\newblock Naturalistic follow-up after a trial of medications for opioid use
  disorder: Medication status, opioid use, and relapse.
\newblock \emph{Journal of substance abuse treatment}, 131:\penalty0 108447,
  2021.

\bibitem[Hamilton(1960)]{hamilton1960hamilton}
Max Hamilton.
\newblock The hamilton depression scale—accelerator or break on
  antidepressant drug discovery.
\newblock \emph{Psychiatry}, 23:\penalty0 56--62, 1960.

\bibitem[Hejazi et~al.(2020)Hejazi, Rudolph, van~der Laan, and
  D{\'\i}az]{hejazi2020nonparametric}
Nima~S Hejazi, Kara~E Rudolph, Mark~J van~der Laan, and Iv{\'a}n D{\'\i}az.
\newblock Nonparametric causal mediation analysis for stochastic interventional
  (in) direct effects.
\newblock \emph{arXiv preprint arXiv:2009.06203}, 2020.

\bibitem[Hser et~al.(2016)Hser, Evans, Huang, Weiss, Saxon, Carroll, Woody,
  Liu, Wakim, Matthews, et~al.]{hser2016long}
Yih-Ing Hser, Elizabeth Evans, David Huang, Robert Weiss, Andrew Saxon,
  Kathleen~M Carroll, George Woody, David Liu, Paul Wakim, Abigail~G Matthews,
  et~al.
\newblock Long-term outcomes after randomization to buprenorphine/naloxone
  versus methadone in a multi-site trial.
\newblock \emph{Addiction}, 111\penalty0 (4):\penalty0 695--705, 2016.

\bibitem[Hser et~al.(2017)Hser, Huang, Saxon, Woody, Moskowitz, Matthews, and
  Ling]{hser2017distinctive}
Yih-Ing Hser, David Huang, Andrew~J Saxon, George Woody, Andrew~L Moskowitz,
  Abigail~G Matthews, and Walter Ling.
\newblock Distinctive trajectories of opioid use over an extended follow-up of
  patients in a multi-site trial on buprenorphine+ naloxone and methadone.
\newblock \emph{Journal of addiction medicine}, 11\penalty0 (1):\penalty0 63,
  2017.

\bibitem[Huang and Yang(2017)]{huang2017causal}
Yen-Tsung Huang and Hwai-I Yang.
\newblock Causal mediation analysis of survival outcome with multiple
  mediators.
\newblock \emph{Epidemiology (Cambridge, Mass.)}, 28\penalty0 (3):\penalty0
  370, 2017.

\bibitem[Kennedy et~al.(2021)Kennedy, Balakrishnan, and
  Wasserman]{kennedy2021semiparametric}
Edward~H Kennedy, Sivaraman Balakrishnan, and Larry Wasserman.
\newblock Semiparametric counterfactual density estimation.
\newblock \emph{arXiv preprint arXiv:2102.12034}, 2021.

\bibitem[Klaassen(1987)]{klaassen1987consistent}
Chris~AJ Klaassen.
\newblock Consistent estimation of the influence function of locally
  asymptotically linear estimators.
\newblock \emph{The Annals of Statistics}, 15\penalty0 (4):\penalty0
  1548--1562, 1987.

\bibitem[Lee et~al.(2018)Lee, Nunes~Jr, Novo, Bachrach, Bailey, Bhatt, Farkas,
  Fishman, Gauthier, Hodgkins, et~al.]{lee2018comparative}
Joshua~D Lee, Edward~V Nunes~Jr, Patricia Novo, Ken Bachrach, Genie~L Bailey,
  Snehal Bhatt, Sarah Farkas, Marc Fishman, Phoebe Gauthier, Candace~C
  Hodgkins, et~al.
\newblock Comparative effectiveness of extended-release naltrexone versus
  buprenorphine-naloxone for opioid relapse prevention (x: Bot): a multicentre,
  open-label, randomised controlled trial.
\newblock \emph{The Lancet}, 391\penalty0 (10118):\penalty0 309--318, 2018.

\bibitem[Lin et~al.(2017)Lin, Young, Logan, and VanderWeele]{lin2017mediation}
Sheng-Hsuan Lin, Jessica~G Young, Roger Logan, and Tyler~J VanderWeele.
\newblock Mediation analysis for a survival outcome with time-varying
  exposures, mediators, and confounders.
\newblock \emph{Statistics in medicine}, 36\penalty0 (26):\penalty0 4153--4166,
  2017.

\bibitem[Luedtke et~al.(2017)Luedtke, Sofrygin, van~der Laan, and
  Carone]{luedtke2017sequential}
Alexander~R Luedtke, Oleg Sofrygin, Mark~J van~der Laan, and Marco Carone.
\newblock Sequential double robustness in right-censored longitudinal models.
\newblock \emph{arXiv preprint arXiv:1705.02459}, 2017.

\bibitem[Miles(2021)]{miles2021jsm}
Caleb Miles, editor.
\newblock \emph{When Randomized Interventional Indirect Effects Tell Stories
  About Mediated Effects (and When They Don’t)}, August 2021. ASA.
\newblock URL \url{https://calebhmiles.github.io/files/2021_JSM-talk.pdf}.

\bibitem[Miles et~al.(2015)Miles, Kanki, Meloni, and
  Tchetgen]{miles2015partial}
Caleb~H Miles, Phyllis Kanki, Seema Meloni, and Eric J~Tchetgen Tchetgen.
\newblock On partial identification of the pure direct effect.
\newblock \emph{arXiv preprint arXiv:1509.01652}, 2015.

\bibitem[Mittinty and Vansteelandt(2020)]{mittinty2020longitudinal}
Murthy~N Mittinty and Stijn Vansteelandt.
\newblock Longitudinal mediation analysis using natural effect models.
\newblock \emph{American Journal of Epidemiology}, 189\penalty0 (11):\penalty0
  1427--1435, 2020.

\bibitem[Moreno-Betancur and Carlin(2018)]{moreno2018understanding}
Margarita Moreno-Betancur and John~B Carlin.
\newblock Understanding interventional effects: a more natural approach to
  mediation analysis?
\newblock \emph{Epidemiology}, 29\penalty0 (5):\penalty0 614--617, 2018.

\bibitem[Pearl(2000)]{Pearl00}
Judea Pearl.
\newblock \emph{Causality: Models, Reasoning, and Inference}.
\newblock Cambridge University Press, Cambridge, 2000.

\bibitem[Pearl(2001)]{Pearl01}
Judea Pearl.
\newblock Direct \& indirect effects.
\newblock In \emph{Proceedings of the 17th Conference in Uncertainty in
  Artificial Intelligence}, UAI '01, pages 411--420, San Francisco, CA, USA,
  2001. Morgan Kaufmann Publishers Inc.
\newblock ISBN 1-55860-800-1.
\newblock URL \url{http://dl.acm.org/citation.cfm?id=647235.720084}.

\bibitem[Petersen et~al.(2006)Petersen, Sinisi, and van~der
  Laan]{petersen2006estimation}
Maya~L Petersen, Sandra~E Sinisi, and Mark~J van~der Laan.
\newblock Estimation of direct causal effects.
\newblock \emph{Epidemiology}, pages 276--284, 2006.

\bibitem[Robins et~al.(2009)Robins, Li, Tchetgen, and van~der
  Vaart]{robins2009quadratic}
James Robins, Lingling Li, Eric Tchetgen, and Aad~W van~der Vaart.
\newblock Quadratic semiparametric von mises calculus.
\newblock \emph{Metrika}, 69\penalty0 (2-3):\penalty0 227--247, 2009.

\bibitem[Robins and Greenland(1992)]{RobinsGreenland92}
James~M Robins and Sander Greenland.
\newblock Identifiability and exchangeability for direct and indirect effects.
\newblock \emph{Epidemiology}, 3\penalty0 (0):\penalty0 143--155, 1992.

\bibitem[Robins and Richardson(2010)]{robins2010alternative}
James~M Robins and Thomas~S Richardson.
\newblock Alternative graphical causal models and the identification of direct
  effects.
\newblock \emph{Causality and psychopathology: Finding the determinants of
  disorders and their cures}, pages 103--158, 2010.

\bibitem[Robins(2000)]{Robins00}
J.M. Robins.
\newblock Robust estimation in sequentially ignorable missing data and causal
  inference models.
\newblock In \emph{Proceedings of the American Statistical Association}, 2000.

\bibitem[Robins et~al.(1994)Robins, Rotnitzky, and
  Zhao]{Robins&Rotnitzky&Zhao94}
J.M. Robins, A.~Rotnitzky, and L.P. Zhao.
\newblock Estimation of regression coefficients when some regressors are not
  always observed.
\newblock \emph{Journal of the American Statistical Association}, 89\penalty0
  (427):\penalty0 846--866, September 1994.

\bibitem[Rotnitzky et~al.(2017)Rotnitzky, Robins, and
  Babino]{rotnitzky2017multiply}
Andrea Rotnitzky, James Robins, and Lucia Babino.
\newblock On the multiply robust estimation of the mean of the g-functional.
\newblock \emph{arXiv preprint arXiv:1705.08582}, 2017.

\bibitem[Rudolph et~al.(2021{\natexlab{a}})Rudolph, D{\'\i}az, Hejazi, van~der
  Laan, Luo, Shulman, Campbell, Rotrosen, and Nunes]{rudolph2021explaining}
Kara~E Rudolph, Iv{\'a}n D{\'\i}az, Nima~S Hejazi, Mark~J van~der Laan, Sean~X
  Luo, Matisyahu Shulman, Aimee Campbell, John Rotrosen, and Edward~V Nunes.
\newblock Explaining differential effects of medication for opioid use disorder
  using a novel approach incorporating mediating variables.
\newblock \emph{Addiction}, 116\penalty0 (8):\penalty0 2094--2103,
  2021{\natexlab{a}}.

\bibitem[Rudolph et~al.(2021{\natexlab{b}})Rudolph, Gimbrone, and
  D{\'\i}az]{rudolph2021helped}
Kara~E Rudolph, Catherine Gimbrone, and Iv{\'a}n D{\'\i}az.
\newblock Helped into harm: Mediation of a housing voucher intervention on
  mental health and substance use in boys.
\newblock \emph{Epidemiology}, 32\penalty0 (3):\penalty0 336--346,
  2021{\natexlab{b}}.

\bibitem[SAMHSA(2021)]{tip2021}
SAMHSA.
\newblock Medications for opioid use disorder for healthcare and addiction
  professionals, policymakers, patients, and families: treatment improvement
  protocol tip 63.
\newblock 2021.
\newblock URL
  \url{https://store.samhsa.gov/sites/default/files/SAMHSA_Digital_Download/PEP21-02-01-002.pdf}.

\bibitem[Sobell and Sobell(1995)]{sobell1995alcohol}
LC~Sobell and MB~Sobell.
\newblock Alcohol timeline followback users’ manual.
\newblock \emph{Toronto, Canada: Addiction Research Foundation}, 1995.

\bibitem[Solli et~al.(2018)Solli, Latif, Opheim, Krajci, Sharma-Haase, Benth,
  Tanum, and Kunoe]{solli2018effectiveness}
Kristin~Klemmetsby Solli, Zill-e-Huma Latif, Arild Opheim, Peter Krajci, Kamni
  Sharma-Haase, J{\=u}rat{\.e}~{\v{S}}altyt{\.e} Benth, Lars Tanum, and Nikolaj
  Kunoe.
\newblock Effectiveness, safety and feasibility of extended-release naltrexone
  for opioid dependence: a 9-month follow-up to a 3-month randomized trial.
\newblock \emph{Addiction}, 113\penalty0 (10):\penalty0 1840--1849, 2018.

\bibitem[Tchetgen and Phiri(2014)]{tchetgen2014bounds}
Eric J.~Tchetgen Tchetgen and Kelesitse Phiri.
\newblock Bounds for pure direct effect.
\newblock \emph{Epidemiology (Cambridge, Mass.)}, 25\penalty0 (5):\penalty0
  775, 2014.

\bibitem[van~der Laan and Petersen(2008)]{van2008direct}
Mark~J van~der Laan and Maya~L Petersen.
\newblock Direct effect models.
\newblock \emph{The international journal of biostatistics}, 4\penalty0 (1),
  2008.

\bibitem[{van der Laan} and Robins(2003)]{vanderLaan2003}
Mark~J {van der Laan} and James~M Robins.
\newblock \emph{Unified Methods for Censored Longitudinal Data and Causality}.
\newblock Springer, New York, 2003.

\bibitem[{van der Laan} and Rose(2011)]{vanderLaanRose11}
Mark~J {van der Laan} and Sherri Rose.
\newblock \emph{Targeted Learning: Causal Inference for Observational and
  Experimental Data}.
\newblock Springer, New York, 2011.

\bibitem[{van der Laan} and Rose(2018)]{vanderLaanRose18}
Mark~J {van der Laan} and Sherri Rose.
\newblock \emph{Targeted Learning in Data Science: Causal Inference for Complex
  longitudinal Studies}.
\newblock Springer, New York, 2018.

\bibitem[{van der Laan} and Rubin(2006)]{vdl2006targeted}
Mark~J {van der Laan} and Daniel Rubin.
\newblock Targeted maximum likelihood learning.
\newblock \emph{The International Journal of Biostatistics}, 2\penalty0 (1),
  2006.

\bibitem[van~der Laan et~al.(2007)van~der Laan, Polley, and
  Hubbard]{vanderLaanPolleyHubbard07}
M.J. van~der Laan, E.~Polley, and A.~Hubbard.
\newblock Super learner.
\newblock \emph{Statistical Applications in Genetics \& Molecular Biology},
  6\penalty0 (25):\penalty0 Article 25, 2007.

\bibitem[van~der Vaart(1998)]{vanderVaart98}
A.~W. van~der Vaart.
\newblock \emph{Asymptotic Statistics}.
\newblock Cambridge University Press, 1998.

\bibitem[VanderWeele(2009)]{vanderweele2009mediation}
Tyler~J VanderWeele.
\newblock Mediation and mechanism.
\newblock \emph{European journal of epidemiology}, 24\penalty0 (5):\penalty0
  217--224, 2009.

\bibitem[VanderWeele and Tchetgen(2017)]{vanderweele2017mediation}
Tyler~J VanderWeele and Eric J~Tchetgen Tchetgen.
\newblock Mediation analysis with time varying exposures and mediators.
\newblock \emph{Journal of the Royal Statistical Society. Series B, Statistical
  Methodology}, 79\penalty0 (3):\penalty0 917, 2017.

\bibitem[VanderWeele et~al.(2014)VanderWeele, Vansteelandt, and
  Robins]{vanderweele2014effect}
Tyler~J VanderWeele, Stijn Vansteelandt, and James~M Robins.
\newblock Effect decomposition in the presence of an exposure-induced
  mediator-outcome confounder.
\newblock \emph{Epidemiology (Cambridge, Mass.)}, 25\penalty0 (2):\penalty0
  300, 2014.

\bibitem[Vansteelandt and Daniel(2017)]{vansteelandt2017interventional}
Stijn Vansteelandt and Rhian~M Daniel.
\newblock Interventional effects for mediation analysis with multiple
  mediators.
\newblock \emph{Epidemiology (Cambridge, Mass.)}, 28\penalty0 (2):\penalty0
  258, 2017.

\bibitem[Vansteelandt et~al.(2019)Vansteelandt, Linder, Vandenberghe, Steen,
  and Madsen]{vansteelandt2019mediation}
Stijn Vansteelandt, Martin Linder, Sjouke Vandenberghe, Johan Steen, and Jesper
  Madsen.
\newblock Mediation analysis of time-to-event endpoints accounting for
  repeatedly measured mediators subject to time-varying confounding.
\newblock \emph{Statistics in medicine}, 38\penalty0 (24):\penalty0 4828--4840,
  2019.

\bibitem[{von Mises}(1947)]{mises1947asymptotic}
R~{von Mises}.
\newblock On the asymptotic distribution of differentiable statistical
  functions.
\newblock \emph{The annals of mathematical statistics}, 18\penalty0
  (3):\penalty0 309--348, 1947.

\bibitem[Wager and Walther(2015)]{wager2015adaptive}
Stefan Wager and Guenther Walther.
\newblock Adaptive concentration of regression trees, with application to
  random forests.
\newblock \emph{arXiv preprint arXiv:1503.06388}, 2015.

\bibitem[Weiss et~al.(2011)Weiss, Potter, Fiellin, Byrne, Connery, Dickinson,
  Gardin, Griffin, Gourevitch, Haller, et~al.]{weiss2011adjunctive}
Roger~D Weiss, Jennifer~Sharpe Potter, David~A Fiellin, Marilyn Byrne, Hilary~S
  Connery, William Dickinson, John Gardin, Margaret~L Griffin, Marc~N
  Gourevitch, Deborah~L Haller, et~al.
\newblock Adjunctive counseling during brief and extended
  buprenorphine-naloxone treatment for prescription opioid dependence: a
  2-phase randomized controlled trial.
\newblock \emph{Archives of general psychiatry}, 68\penalty0 (12):\penalty0
  1238--1246, 2011.

\bibitem[Weiss et~al.(2015)Weiss, Potter, Griffin, Provost, Fitzmaurice,
  McDermott, Srisarajivakul, Dodd, Dreifuss, McHugh, et~al.]{weiss2015long}
Roger~D Weiss, Jennifer~Sharpe Potter, Margaret~L Griffin, Scott~E Provost,
  Garrett~M Fitzmaurice, Katherine~A McDermott, Emily~N Srisarajivakul,
  Dorian~R Dodd, Jessica~A Dreifuss, R~Kathryn McHugh, et~al.
\newblock Long-term outcomes from the national drug abuse treatment clinical
  trials network prescription opioid addiction treatment study.
\newblock \emph{Drug and alcohol dependence}, 150:\penalty0 112--119, 2015.

\bibitem[Zheng and van~der Laan(2017)]{zheng2017longitudinal}
Wenjing Zheng and Mark van~der Laan.
\newblock Longitudinal mediation analysis with time-varying mediators and
  exposures, with application to survival outcomes.
\newblock \emph{Journal of causal inference}, 5\penalty0 (2), 2017.

\bibitem[Zheng and van~der Laan(2011)]{zheng2011cross}
Wenjing Zheng and Mark~J van~der Laan.
\newblock Cross-validated targeted minimum-loss-based estimation.
\newblock In \emph{Targeted Learning}, pages 459--474. Springer, 2011.

\end{thebibliography}


\begin{thebibliography}{6}
\providecommand{\natexlab}[1]{#1}
\providecommand{\url}[1]{\texttt{#1}}
\expandafter\ifx\csname urlstyle\endcsname\relax
  \providecommand{\doi}[1]{doi: #1}\else
  \providecommand{\doi}{doi: \begingroup \urlstyle{rm}\Url}\fi

\bibitem[Bang and Robins(2005)]{Bang05}
Heejung Bang and James~M Robins.
\newblock Doubly robust estimation in missing data and causal inference models.
\newblock \emph{Biometrics}, 61\penalty0 (4):\penalty0 962--973, 2005.

\bibitem[D{\'\i}az et~al.(2021)D{\'\i}az, Williams, Hoffman, and
  Schenck]{diaz2021nonparametricmtp}
Iv{\'a}n D{\'\i}az, Nicholas Williams, Katherine~L Hoffman, and Edward~J
  Schenck.
\newblock Nonparametric causal effects based on longitudinal modified treatment
  policies.
\newblock \emph{Journal of the American Statistical Association}, pages 1--16,
  2021.

\bibitem[Luedtke et~al.(2017)Luedtke, Sofrygin, van~der Laan, and
  Carone]{luedtke2017sequential}
Alexander~R Luedtke, Oleg Sofrygin, Mark~J van~der Laan, and Marco Carone.
\newblock Sequential double robustness in right-censored longitudinal models.
\newblock \emph{arXiv preprint arXiv:1705.02459}, 2017.

\bibitem[Robins(1986)]{Robins86}
James~M Robins.
\newblock A new approach to causal inference in mortality studies with
  sustained exposure periods - application to control of the healthy worker
  survivor effect.
\newblock \emph{Mathematical Modelling}, 7:\penalty0 1393--1512, 1986.

\bibitem[Robins(1999)]{Robins99}
James~M Robins.
\newblock Robust estimation in sequentially ignorable missing data and causal
  inference models.
\newblock \emph{Proc of the Am Stat Assoc Sec on Bayes Stat Sci}, pages 6--10,
  1999.

\bibitem[van~der Laan and Gruber(2012)]{van2012targeted}
Mark~J van~der Laan and Susan Gruber.
\newblock Targeted minimum loss based estimation of causal effects of multiple
  time point interventions.
\newblock \emph{The international journal of biostatistics}, 8\penalty0 (1),
  2012.

\end{thebibliography}
